\documentclass[reprint,secnumarabic,amssymb, nobibnotes]{jfm}

\usepackage{color}

\usepackage{amssymb,amsmath}
\usepackage{graphicx}
\usepackage{natbib}
\usepackage{wasysym}
\usepackage{cases}

\def\REV#1{{\textcolor{black}{ #1}}}

\begin{document}

\title{Turbulent pair dispersion as a ballistic cascade phenomenology.}

\author{Micka\"el Bourgoin}
\affiliation{Laboratoire des \'Ecoulements G\'eophysiques et Industriels, CNRS/UJF/G-INP, UMR 5519, Universit\'e de Grenoble, BP53, 38041, Grenoble, France}

\maketitle

\begin{abstract}
Since the pioneering work of Richardson in 1926, later refined by Batchelor and Obukhov in 1950, it is predicted that the rate of separation of pairs of fluid elements in turbulent flows with initial separation at inertial scales, grows ballistically first (Batchelor regime), before undergoing a transition towards a super-diffusive regime where the mean-square separation grows as $t^3$ (Richardson regime). Richardson empirically interpreted this super-diffusive regime in terms of a non-Fickian process with a scale dependent diffusion coefficient (the celebrated Richardson's ``4/3rd'' law). However, the actual physical mechanism at the origin of such a scale dependent diffusion coefficient remains unclear. The present article proposes a simple physical phenomenology for the time evolution of the mean square relative separation in turbulent flows, based on a scale dependent \emph{ballistic} scenario rather than a scale dependent \emph{diffusive}. It is shown that this phenomenology accurately retrieves most of the known features of relative dispersion for particles mean square separation ; among others : (i) it is quantitatively consistent with most recent numerical simulations and experiments for mean square separation between particles (both for the short term Batchelor regime and the long term Richardson regime, and for all initial separations at inertial scales), (ii) it gives a simple physical explanation of the origin of the super diffusive $t^3$ Richardson regime which naturally builts itself as an iterative process of elementary short-term-scale-dependent ballistic steps, (iii) it shows that the Richardson constant is directly related to the Kolmogorov constant (and eventually to a ballistic persistence parameter) and (iv) in a further extension of the phenomenology, taking into account third order corrections, it robustly describes the temporal asymmetry between forward and backward dispersion, with an explicit connection to the cascade of energy flux across scales. An important aspect of this phenomenology is that it simply and robustly connects long term super-diffusive features to elementary short term mechanisms, and at the same time it connects basic Lagrangian features of turbulent relative dispersion (both at short and long times) to basic Eulerian features of the turbulent field : second order Eulerian statistics control the growth of separation (both at short and long times) while third order Eulerian statistics control the temporal asymmetry of the dispersion process, which can then be directly identified as the signature of the energy cascade and associated to well known exact results as the Karman-Howarth-Monin relation.
\end{abstract}

\section{Introduction}
Molecules in a quiescent fluid tend to spread due to molecular diffusion. If we consider a small spherical patch of tagged molecules, this results in an isotropic and homogeneous growth of the patch. At a microscopic level this expansion is due to random uncorrelated collisions induced by the thermal agitation of the molecules. At a macroscopic level this mechanism results in a Fickian diffusion process where the local concentration $C$ of tagged molecules diffuses according to the simple equation $\partial C/\partial t=K \Delta C$, where $K$ is the molecular diffusivity, with units [m$^2\cdot$s$^{-1}$]. In elementary kinetic gas theory, the connection between microscopic and macroscopic descriptions is for instance given by the relation $K\propto lv_T$ (with $l$ a characteristic correlation length of particles trajectories, typically given by the mean free path and $v_T$ the thermal agitation velocity of the molecules). A fundamental property of such a Fickian process concerns the linear growth with time $t$ of the mean square separation $< \vec{D}^2>\propto Kt$ between any two molecules in the patch, what is generally referred to as \emph{normal diffusion}. It is well-known that normal molecular diffusion alone is very inefficient to mix and disperse usual species (for instance, molecular diffusivity of carbon dioxyde in air is 16$\cdot10^{-6}$m$^2\cdot$s$^{-1}$, meaning that molecules separate at a rate of only a few millimeters per second). 

A usual way to enhance mixing and dispersion consists in stirring the fluid in order to generate large scale uncorrelated turbulent structures, which will act in a similar way (\emph{i.e.} normally diffusive) as molecular diffusion, but with an enhanced diffusion coefficient $K_{\textrm{turb}}\propto L \sigma$ with $L$ the turbulence correlation length scale and $\sigma$ the turbulent fluctuating velocity (standard deviation of the turbulent velocity field). In the context of atmospheric dipersion for instance, the turbulent correlation length is typically of the order of hundreds of meters (let  us take 100~m as an order of magnitude) with velocity fluctuations typically of the order of meters per second in normal conditions (let us take 1~m/s as an order of magnitude), leading to a turbulent diffusivity coefficient $K_{\textrm{turb}}$ of the order of 30~m$^2\cdot$s$^{-1}$ (meaning that fluid particles separate at a rate of several meters per second), hence many orders of magnitude larger than molecular diffusion. The efficiency of turbulent diffusion therefore relies on the capacity of a substance to spread thanks to the uncorrelated motion of large scale turbulent eddies. However, if we consider the dispersion of a patch initially much smaller than the turbulent correlation scale $L$ (for instance a patch with an initial dimension within the inertial range of the carrier turbulence, hence much smaller than the energy injection scale $L$, and larger than the dissipation scale $\eta$), another mechanism is necessary to allow the patch to grow first at sufficiently large scales to eventually undergo the effect of uncorrelated turbulent diffusion. Such an inertial scale mechanism is ensured by the \emph{super-diffusive} nature of turbulence at inertial scales. Processes where the mean square separation grows faster than in normal diffusion (\emph{i.e.} $< \vec{D}^2>\propto t^\alpha$, with $\alpha>1$) are called \emph{super-diffusive}. 
\begin{figure}
\center
\includegraphics[height=7cm]{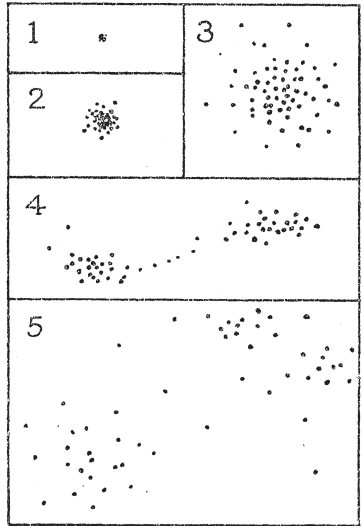}
\caption{Qualitative illustration of the non-normal dispersion of a dense cluster of particles as proposed in Richardson's original 1926 article~[\cite{bib:richardson1926}], from which the figure is taken.}\label{fig:ballisticRichardson}
\end{figure}

The super-diffusive nature of turbulence and its major consequences in terms of enhanced mixing was first emphasized by Lewis Richardson in his seminal 1926 article~(see \cite{bib:richardson1926}). Concerning the dispersal of an initial point-charge of particles in the atmosphere, Richardson already noted in 1926 that ``\emph{a small dense cluster of marked molecules, represented by the dot in figure 1(1) which, by molecular diffusion alone, would spread through the successive spherical clusters shown in figures 1(2) and 1(3), actually seldom passes through the large spherical stage 1(3), because it is first sheared into two detached clusters as suggested in figure 1(4). These are carried far from one another, and are likely to be again torn into smaller pieces as in figure 1(5)}''. This qualitative description by Richardson shows how turbulence acts to super-diffusively separate particles initially packed in a small patch, in order to create sufficiently large separations where the uncorrelated motion of turbulent eddies eventually disperses particles at large scales.

Richardson gave an interpretation of turbulent super-diffusion in terms of a non-Fickian process which could be locally modeled as a normal diffusion process, but with a scale dependent diffusion coefficient which depends on particle separation $D$, according to the celebrated Richardson's 4/3rd law : $K(D)\propto D^{4/3}$. Besides, Richardson showed that this non-Fickian diffusion resulted in a cubic super-diffusive growth of the mean square separation of pairs of particles according to the law  $\left<D^2\right>=g\epsilon t^3$, where $\epsilon$ is the turbulent energy dissipation rate and $g$ a universal constant since known as the Richardson constant. In the framework of Kolmogorov 1941 phenomenology of turbulence~(\cite{bib:K41}, hereafter referred as K41) the $t^3$ dependency can be  understood as a simple dimensional constraint. K41 states indeed that for sufficiently large Reynolds number, the only relevant  physical parameter for the dynamics of turbulence at inertial scales is the average energy dissipation rate per unit mass $\epsilon$ (with dimensions~[m$^2\cdot$s$^{-3}$]): $D^2\propto \epsilon t^3$ is then the only dimensionally consistent relation if initial separation is ignored.

Richardson's work was later refined by Batchelor and Obukhov in the 1950s~(see \cite{bib:batchelor1950}), who pointed that while the loss of memory of initial separation is a reasonable assumption for the long-term dispersion, initial separation must play a role in the short-term. They showed that the rate of separation of pairs of fluid elements in turbulent flows with initial separation $\vec{D}_0$ at inertial scales ($\eta \ll D_0 \ll L$) must obey the following scalings :

\begin{subnumcases}{R^2=\left<\left(\vec{D}-\vec{D}_0\right)^2\right>=}
		S_2(\vec{D}_0)t^2	&	\text{if $t<t_0$} \label{eq:disp1a}\\
		g\epsilon t^3.	&	\text{if $t>t_0$}\label{eq:disp1b}
\end{subnumcases}
with $S_2(\vec{r})=\left<\left| \delta_{\vec{r}} \vec{u}\right|^2\right>$ the full second order Eulerian structure function of the velocity field (with $\delta_{\vec{r}}\vec{u}$ the increment between two points separated by a vector $\vec{r}$ of the eulerian velocity field of the flow ; note that homogeneity is assumed, so that velocity increment only depends on the separation vector) and $t_0$ a characteristic time scale of the particles motion at scale $D_0$. \REV{In K41 framework, inertial scalings for $S_2$ and $t_0$ are : $S_2(\vec{D}_0)=\frac{11}{3}C_2 \epsilon^{2/3}D_0^{2/3}$ (where local isotropy is also assumed so that $S_2(\vec{r})$ only depends on the norm of the separation vector $r=\sqrt{|\vec{r}|^2}$, and $C_2$ is a universal constant known as the Kolmogorov constant, with a well accepted value $C_2\simeq 2.1$ for 3D homogeneous and isotropic turbulence, see \cite{bib:sreenivasan1995_PoF}) and $t_0\propto \epsilon^{-1/3}D_0^{2/3}$} ($t_0$ then represents the eddy turnover time at scale $D_0$). 
Formally speaking, the initial ballistic regime~(eq.~\ref{eq:disp1a}) is nothing but the leading term of the Taylor expansion for the mean square pair separation at short times, expressed in terms of the initial mean square relative velocity between particles~(\cite{bib:batchelor1950,bib:ouellette2006b_NJP}). Note that such a ballistic Taylor expansion is a general and purely kinematic relation valid for any early dispersion process and is not limited to the case of turbulence. Specificities of turbulence only appear when expliciting the form of the structure function $S_2$ at inertial scales. 
This short-term ballistic regime has been shown to be accurately and robustly followed in experiments of turbulent relative pair dispersion within the inertial scales of 3D-turbulence~(see for instance~\cite{bib:bourgoin2006_Science}). 

For times exceeding $t_0$, a transition is expected towards an enhanced dispersion regime, cubic in time and independent of initial separaion, as originally predicted by Richardson. \REV{The experimental observation of this regime remains however very elusive. A convincing $t^3$ regime has been reported for 2D-turbulence experiments by \cite{bib:jullien1999_PRL}. Concerning 3D-turbulence experiments, \cite{bib:ott2000_JFM} reported a cubic regime, but they needed to introduce a time-shift in the cubic expression~\ref{eq:disp1b} ($<(D-D_0)^2>\propto \epsilon (t-t_0)^3$, with $t_0$ a negative virtual time origin depending on initial separation) to fit accurately their experimental data.} 

The Richardson constant $g$ in eq.~(\ref{eq:disp1b}) is one of the most fundamental constants in turbulence. It plays a major role in turbulent dispersion and mixing processes. However, in spite of its importance, it is only recently that a robust estimate for $g$ started to emerge in the litterature~(\cite{bib:sawford2001,bib:salazar2009_ARFM}). This is probably related to the difficulty to observe experimentally Richardson's superdiffusion. Until recently, best estimates for $g$ still spanned several orders of magnitude. Most recent high resolution direct numerical simulations seem to point toward a robust estimate of $g\sim 0.5-0.6$~(\cite{bib:boffetta2002_PRE,bib:biferale2005_PoF,bib:bitane2012_PRE}), in agreement with the experiments by~\cite{bib:ott2000_JFM}, where $g$ was found around 0.55.

As already mentioned, in his seminal 1926 article~(\cite{bib:richardson1926}), Richardson empirically related such a superdiffusive regime to a non-Fickian process, with a local diffusivity coefficient $K$ which depends on the probed spatial scale $D$: $K(D)\propto D^{4/3}$. Further refinements in the framework of K41 phenomenolgy extended Richardson's non-Fickian phenomenology by considering also a time scale dependency of the local diffusivity coefficient~(\cite{bib:klafter1987_PRE}) such that $K(D,\tau)= k_0 \epsilon^\gamma D^\alpha\tau^\beta$ (with the dimensional constraints that $3\alpha+2\beta=4$ and $\gamma=1-\alpha/2$). Such processes also lead to a $t^3$ regime for the mean square separation (the Richardson constant $g$ is then directly related to $k_0$). However no clear physical interpretation for the  origin of such a time/scale dependency of the local diffusivity is still known, \REV{even if the problem of relative dispersion has been extensively studied in the last decades, both for 3D and 2D-turbulence, and from theoretical, numerical and experimental points of view. A complete review of the turbulent relative dispersion problem goes beyond the scope of the present article, and I suggest to the reader to refer to two important reviews on the question~(\cite{bib:sawford2001, bib:salazar2009_ARFM}) for a detailed insight into this topic. In this rich context the present work is however in the direct line of some previous studies, seeking for possible physical interpretations to explain Richardson super-diffusion and for possible connections with K41 phenomenology (beyond purely dimensional considerations), among wihch I would like to stress some important contributions.}

\REV{Grossman \& Procaccia (\cite{bib:grossmann1984_PRA,bib:grossmann1990}) proposed a model for pair dispersion relying on a mean field approach of Navier-Stokes equation and a phenomenological closure assumption which allowed them to relate the Richardson constant to the Kolmogorov constant $C_2$ as $g=(\frac{22}{9}C_2)^{3/2}$ without any adjustable parameter. In spite of the elegance of this model, the predicted value for $g\simeq 12$ (taking the well accepted value for $C_2\simeq 2.1$) remains about 20 times larger compared to the current reference value for $g\simeq0.5-0.6$.}

\REV{More classical approaches of pair dispersion tend simply to follow Richardson's original idea of a scale dependent diffusive process. However,~\cite{bib:sokolov1999b_PRE} emphasized that in flows where two point velocities statistics and local correlation time scales follow K41 dimensional constraints, a local ballistic description may be more relevant than a local diffusive approach, what is also supported by experiments and numerical studies where, as already mentioned, the short term ballisitic behavior is indeed robustly observed. \cite{bib:sokolov2000_PRE} then proposed a heuristic model for pair dispersion including a persistence parameter weighting the relative importance of diffusive versus ballistic processes. This model, based on a one dimensional L\'evy-walk description of pair dispersion, considers a succession of simple ballistic separations between random turning points, with scale dependent relative velocities and a persistence parameter describing the probability of turning points. The model was shown by the authors to be in good qualitative agreement with experimental observations of Richardson dispersion in  2D-turbulence by~\cite{bib:jullien1999_PRL}, confirming the leading role of short term ballistic events in the overall turbulent super-diffusion process.} 

\REV{More recently, \cite{bib:goto2004_NJP} proposed an simple similar model (known as GV04 model), also higlighting the role of ballisitic motion between turning points (identified here as the zero-accleration points of the flow field) where particles separation undergoes sudden growing ``bursts''. In the GV04 model the dispersion process is described as a step by step process where the separation $D_k$ at step $k$ grows to $D_{k+1}=\xi D_k$ (with $\xi$ a prescribed growth parameter) after a scale depend waiting time  $T_\xi(D_k)$ depending on the number of acceleration stagnation points in the flow. When K41 scalings are considered for the time scale $T_\xi$ this model naturally builds a long term $t^3$ Richardson regime, where Richardson's constant is simply related to the parameter $\xi$ and to the number of acceleration stagnation points. In a further refinement of this model, \cite{bib:faber2009_PoF} added the possibility for particles to converge to smaller scales. The refined model, applied to two-dimensional turbulence, was shown to reproduce the temporal asymmetry of particle forward \emph{vs} backward dispersion in 2D-turbulence DNS. The concept of backward dispersion is crucial for turbulent mixing and passive scalar studies~(\cite{bib:salazar2009_ARFM}). The temporal asymmetry of pair dispersion, was only recently pointed by~\cite{bib:sawford2005_PoF} who showed that backward dispersion in Lagrangian stochastic models and three-dimensional DNS operates at a significantly faster rate compared to forward dispersion. Recent experiments and simulations by~\cite{bib:berg2006_PRE} confirm this trend with a ratio of backward to forward Richardson constants of the order of 2 for 3D-turbulence. Insterestingly, numerical simulations of 2D-turbulence in the inverse cascade regime by Faber \& Vassilicos present the opposite asymmetry, 2D forward dispersion operating faster than 2D backward dispersion. The origin of this asymmetry remains unclear (and will also be addressed in the final section of this article). \cite{bib:sawford2005_PoF} pointed the importance of odd moments of two points velocity statistics while~\cite{bib:berg2006_PRE} suggested a explanation based on the strain-tensor eigenvalues, which was however shown by~\cite{bib:faber2009_PoF} to be insufficient to explain the difference between 2D and 3D-turbulence, hence emphasizing the possible role of energy flux accross scales, which goes from large to small scales in 3D-turbulence and from small to large scale in 2D-turbulence inverse cascade.
} 

\REV{In a different spirit, the dominant role of ``burst'' events and waiting times has also been recently emphasized by~\cite{bib:rast2011_PRL} who investigated the pair dispersion problem in a simplified point-vortex flow model sharing some common properties with actual turbulent flows (\cite{bib:rast2009_PRE}). Their study points toward a leading role of statistics of delay times, corresponding to periods during which particles in a pair remains close to each other before they separate.}

\REV{The work presented in the this article builds on these previous studies, which emphasized the role of local ballisitic events and local  correlation times. Let me finish this introdcution by mentioning that the ballisitic approach proposed in the present article (and detailed in the next section) is similar to a recent publication by~(\cite{bib:thalabard2014_JFM}), but it was derived independently. Besides as discussed further below, although they are similar, the approaches are exploited in different ways.}



\section{A ballistic cascade phenomenology of pair dispersion}
\begin{figure}
\center
\includegraphics[height=7cm]{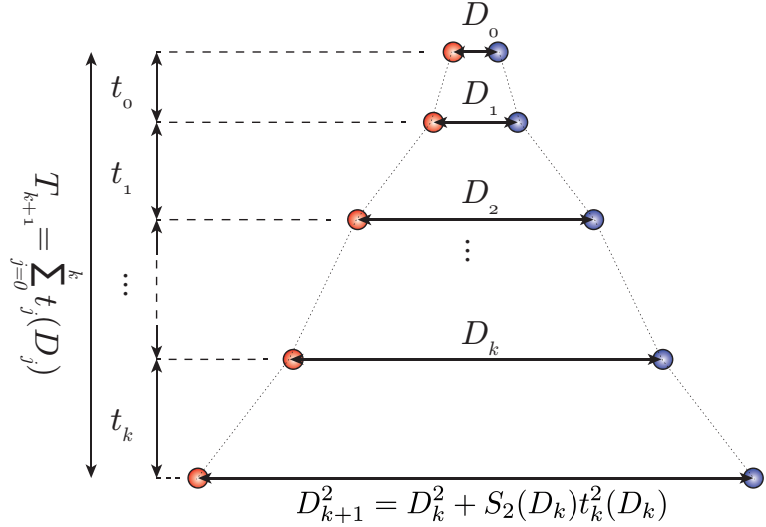}
\caption{Illustration of the iterative ballistic cascade for the relative separation of two particles initially separated by a distance $D_0$: at each iteration step $k$, the mean square separation between particles grows ballistically from $D_k^2$ to $D_{k+1}^2$ with a growth rate $S_2(D_k)$, during a time lag $t_k$. Depending on the physical process at the origin of the local ballistic behavior, both $S_2$ and $t_k$ may eventually depend on the local scale $D_k$. The overall time required to reach the mean square separation $D_k^2$ at the iteration number $k$ is $T_k=\sum_{j=0}^{k-1}t_j(D_j)$.}\label{fig:ballisticCascade}
\end{figure}

I propose here a very simple physical phenomenology for the Richardson super-diffusivity in turbulence, which shares some analogies with the approaches by~\cite{bib:sokolov2000_PRE} and~\cite{bib:faber2009_PoF}. 
The main idea behind the dispersive process proposed here is that of an iterative ballistic mechanism, as illustrated in figure~\ref{fig:ballisticCascade}. It is based on the simple idea that if a set of particle pairs with a given initial mean square separation $D_0^2$ starts to disperse ballistically, with a separation rate $S_2(D_0)$ over a given period $t_0$ after which the mean square separation has grown to $D_1^2=D_0^2+S_2(D_0)t_0^2$ (following the elementary short term ballisitic regime, as given by eq~\ref{eq:disp1a}), instead of considering for $t>t_0$ a sudden transition towards an enhanced cubic dispersion regime (as in eq.~\ref{eq:disp1b}), the same short-term elementary ballistic process can be iterated, but starting from the new mean square separation $D_1^2$, hence with a new separation rate $S_2(\vec{D}_1)$ which operates over a new period of time $t_1$ and so on. Thus, in this scenario the time evolution of particles mean square separation is simply described by the iterative process :

\begin{equation}\label{eq:iterativeScheme}
D_{k+1}^2=D_k^2+S_2(D_k)t_k^2(D_k),
\end{equation}
where $D_k^2=<|\vec{D}_k|^2>$ represents the mean square separation of pairs after the $k^{th}$ iteration step, $t_k(D_k)$ is a scale dependent ``time of flight'' characteristic of the duration of the ballistic motion at step $k+1$. For the case of turbulent flows, $S_2(D_k)$ and $t_k(D_k)$ will be prescribed later by imposing K41 scalings.\\

\subsection{Relevance of the ballisitic phenomenology}\label{sec:relevance}
The present formulation of the ballistic phenomenology addresses the question of the temporal evolution of the mean square separation of an ensemble of pairs ($D_k^2$ in ref~\ref{eq:iterativeScheme} is the mean square separaiton of this ensemble and the growth rates $S_2(D_k)$ and $t_k(D_K)$ have a a statistical meaning only). It is important to stress that although this formulation is limited to the second moment of the statistics of pair separation, it has the benefit of not being just heuristic, but to stand \emph{a priori} on solid statistical grounds, which will be shown later to give \emph{a posteriori} an accurate and quantitative description of the time evolution of pairs mean square separation. First, the short term ballistic statistical relation for the mean square separation in eq.~(\ref{eq:disp1a}), $<(\vec{D}-\vec{D_0})^2>=S_2(D_0)t^2+{\cal{O}}(t^3)$ (which shows that to the leading order, pairs statistics is related to second order velocity increments) is purely kinematic and hence unquestionable. Second, as already discussed in the introduction, this leading role of short term ballistic events in turbulent dispersion statistics and its connection with second order velocity statistics has been emphasized in several theoretical studies and it was accurately observed in high resolution Lagrangian tracking experiments and numerical simulations, with robust scale dependent separation rates and characteristic time scale~(see for instance \cite{bib:bourgoin2006_Science,bib:ouellette2006b_NJP,bib:bitane2012_PRE}). This contrasts with the heuristic approach by~\cite{bib:thalabard2014_JFM}, who applied a similar ballistic scenario to derive iteratively the temporal evolution of separation with a growth rate and a ballistics time defined not in a statistical sense, but for each individual particle pair, 
addressing only in a second stage the statistical implications for an ensemble of such individual realizations (following a continuous-time-random-walk model, similar to the approach by~\cite{bib:sokolov2000_PRE}). As a consequence, although this approach allows to investigate the full statistics of separations built by this process (and not only the mean square separation), it leads to the unrealistic result that the separation growth is governed at leading order by third order velocity increments. 
Since their approach is heuristic, this has probably only minor implications for most of the qualitative results they obtain, but it gives a misleading picture of the relevant physical ingredients at play and it may also lead to misleading conclusions if it is applied to more subtle effects, as for instance the temporal asymmetry of dispersion, for which the actual role of  third order increments may be crucial (this will be discussed in the last section of this article).


Even if the elementary scale dependent ballistic process considered here for the means square separation does indeed rely on solid concepts and observations, several approximations still need to be discussed regarding the proposed iterative approach :
\begin{itemize}
\item To derive the iterative scheme (\ref{eq:iterativeScheme}) I have used the approximation $<D^2>-<D_0^2>=S_2(D_0)t^2$ for the elementary short-term ballisitc process in eq.~(\ref{eq:disp1a}) (which rigorously applies to $<(\vec{D}-\vec{D}_0)^2>$). This is a common approximation, which is equivalent to neglecting the term $<\vec{D}_0\cdot\delta_{\vec{r}}\vec{u}>$~(\cite{bib:batchelor1950}). It is known to hold for statistically homogeneous flows, although it was shown to possibly fail (at very short times) in experiments with non-homogeneous and anisotropic flows~(\cite{bib:ouellette2006b_NJP,bib:salazar2009_ARFM}).\vspace{2mm}
\item A more sever approximation concerns the fact that in the iterative process, the separation rate $S_2(D_k)$ and the ballistic time of flight $t_k(D_k)$ are both estimated assuming a unique value for the separation $D_k$, hence neglecting the statistical fluctuations of pairs square separation, which in the real process 
explore a whole statistical distribution. In the present implementation of the iterative ballisitic model, the separation rate $S_2(D_k)$ and the typical time-scale $t_k(D_k)$ will be simply estimated at the scale corresponding to the mean square separation $D_k^2$. As it will be shown, this apparently na\"ive approximation still gives an accurate quantitative description of the dispersion process when compared to recent numerical simulations of pair dispersion in 3D-turbulence. This approximation, is also supported by the experimental evidence in~\cite{bib:bourgoin2006_Science} that the short term ballistic regime given by eq. (\ref{eq:disp1a}) is robustly observed even if the initial separation is coarsely binned accross inertial scales, and when the growth rate is simply estimated from the mean separation in each bin (note that bins with width up to 100\% of the mean where considered in those experiments, with still an excellent confirmation of the elementary ballistic process~\ref{eq:disp1a}). More quantitatively this approximation can be expected to hold as long as the width, $\delta D^2$, of the distribution of particles square separation remains smaller than the mean square separation $<D^2>$ itself. If $\delta D^2$ is estimated as the standard deviation of particles square separation ($\delta D^2=\sqrt{<(D^2-<D^2>)^2>}$), this condition would be related to the flatness of the separation distribution. 
With this respect, it would be interesting, in a further study, to investigate the ratio $\delta D^2/<D^2>$ and its time evolution based on available experimental and numerical data for the statistical distribution of particles separation, in order to explore more quantitatively the possible range of validity of this approximation.\vspace{2mm}

\item Finally, a concrete implementation of the iterative scheme (\ref{eq:iterativeScheme}) requires the expressions for the scale dependent separation rate $S_2(D_k)$ and the ballistic time of flight $t_k(D_k)$ to be specified. This is discussed in the following sub-section.\\
\end{itemize}


\subsection{Ballistic separation rates and time scales}\label{sec:ballistic}

The separation rate of the elementary ballistic $k^{\mathrm{th}}$ step in (\ref{eq:iterativeScheme}) is simply given by the Eulerian second order structure function at scale $D_k$. In the inertial range of scales of fully developped turbulent flows, under local homogeneity and isotropy assumptions, the second order structure function is known to follow K41 scaling (with negligible intermittent corrections), which in 3D-turbulence can be written as :

\begin{equation}\label{eq:S2}
S_2(\vec{r})=\frac{11}{3}C_2\epsilon^{2/3}r^{2/3},
\end{equation}
with $r=\sqrt{|\vec{r}|^2}$ and where $C_2$ is a universal constant with a well-known value of approximately 2.1; as $C_2$ is analytically related to the Kolmogorov constant $C_K\simeq C_2/4~$ (\cite{bib:sreenivasan1995_PoF}), characterizing the $-5/3$ spectrum of turbulent kinetic energy ($E(k)=C_K\epsilon^{2/3}k^{-5/3}$), we shall equivalently refer to $C_2$ as \emph{the} Kolmogorov constant itself. \REV{Note that $S_2$ here is the total structure function trace and not only the structure function for one specific component of velocity}.\\ 

Concerning the typical time scale $t_0$ for which the elementary ballisitic process in eq.~(\ref{eq:disp1a}) is supposed to hold for a given initial separation $D_0$, Batchelor proposed in his original article~(\cite{bib:batchelor1950}) that it should be taken as the eddy turnover time at the considered scale : $t_0\propto\epsilon^{-1/3}D_0^{2/3}$. Experimental measurements of relative dispersion in highly turbulent flows with Reynolds number up to $R_\lambda\simeq 800$ (\cite{bib:bourgoin2006_Science,bib:ouellette2006b_NJP}) have shown indeed that the time laps for which the ballistic Batchelor regime holds does scale as $\epsilon^{-1/3}D_0^{2/3}$. Recent numerical simulations by~\cite{bib:bitane2012_PRE} have refined the estimation of the characteristic time scale for the ballistic regime. Their simulations show that the duration of the ballistic regime for an initial separation $D_0$ at inertial scales is given by $t_0(D_0)\simeq S_2(D_0)/2\epsilon=\frac{11}{6}C_2\epsilon^{-1/3}D_0^{2/3}$. For times larger than this $t_0$ they observe a transition of the mean square separation  towards a cubic Richardson regime, as in eq.~(\ref{eq:disp1b}). 

The time scale $t_0(D_0)\simeq S_2(D_0)/2\epsilon$ observed in the simulations can be more rigourously discussed by considering the Taylor expansion leading to the short-term ballistic regime in eq.~(\ref{eq:disp1a}) beyond the second order. The third order expansion for $<(\vec{D}-\vec{D}_0)^2>$ (or equivalently for $<D^2>-D_0^2$ under the approximation previously discussed) can be kinematically written as~(\cite{bib:ouellette2006b_NJP,bib:bitane2012_PRE}) :
\begin{equation} \label{eq:3rdOrder}
\left<\vec{D}^2\right>-\left<\vec{D}_0^2\right>=S_2(D_0)t^2+\left<\delta_{\vec{D}_0}\vec{a}\cdot\delta_{\vec{D}_0}\vec{u}\right> t^3+ {\cal{O}}(t^4),
\end{equation}
where $\delta_{\vec{D}_0}\vec{a}$ is the relative pair acceleration and $\delta_{\vec{D}_0}\vec{u}$ the relative pair velocity. The third order coefficient is therefore given by the crossed velocity-acceleration structure funcion $<\delta_{\vec{D}_0}\vec{a}\cdot\delta_{\vec{D}_0}\vec{u}> $ which can be analytically derived from the Navier-Stokes equation~(\cite{bib:risoeReport,bib:hill2006_JoT}) and shown to be equal to $-2\epsilon$ for 3D turbulence, under local stationarity, homogeneity and isotropy conditions (note that $<\delta_{\vec{D}_0}\vec{a}\cdot\delta_{\vec{D}_0}\vec{u}> = -2\epsilon$ is in particular independent of the probed scale $D_0$). The time $t_0(D_0)=S_2(D_0)/2\epsilon$ is therefore exactly the time for which the third order term in eq.~(\ref{eq:3rdOrder}) equals the second order ballistic term (in absolute value). The role of the third order term will be further discussed in section~\ref{sec:asymmetry}, however a brief discussion on the choice of the typical time scale for a purely ballistic (quadratic) dominant regime is still required at this stage. Numerical results by~\cite{bib:bitane2012_PRE} show indeed that for a given initial separation $D_0$, the extent of the short term ballistic regime seems to be accurately given by $S_2(D_0)/2\epsilon$, hence suggesting to use $t_k(D_k)=S_2(D_k)/2\epsilon$ as the typical ``time of flight'' for each elementary ballistic iterative steps in (\ref{eq:iterativeScheme}). This observation raises however the issue that rigorously speaking, considering the negative sign of the third order coefficient $<\delta_{\vec{D}_k}\vec{a}\cdot\delta_{\vec{D}_k}\vec{u}>=-2\epsilon$, the ballistic and the third order term in (\ref{eq:3rdOrder}) cancel exactly for $t=t_k(D_k)$. For times $t\simeq t_k$, negative third order contributions annihilate the initial quadratic growth. In other words, a dominant ballistic mechanism, which requires terms of order 3 and higher to be negligible in the Taylor expansion~(\ref{eq:3rdOrder}) only holds for times $t\ll t_k=S_2(D_k)/2\epsilon$. For the iterative scheme considered here, this indicates that the actual temporal duration of each elementary ballistic step is necessarily much shorter than $t_k(D_k)=S_2(D_k)/2\epsilon$. This points towards the necessity to introduce a ballisitic persistence parameter $\alpha < 1$, defining the actual characteristic ballisitic time of flight at each iteration step $k$ as $t'_k(D_k)=\alpha t_k(D_k)=\alpha S_2(D_k)/2\epsilon$. The condition $\alpha < 1$ (and more realistically $\alpha \ll 1$) ensures that the ballistic term is indeed still dominant at the end of each ballistic iteration of duration $t'_k$ ; we shall refer to this condition as the \emph{ballisitic approximation}. 


\subsection{Explicit formulation of the iterative ballistic model}
We can now explicitly write the complete iterative scheme of the iterative ballistic phenomenology as

\begin{equation}\label{eq:cascade}
D_{k+1}^2=D_k^2+S_2(D_k) t^{\prime 2}_k(D_k) \;\; \textrm{ with }\;\; \left\{
\begin{array}{l}
	S_2(D_k)=C\epsilon^{2/3}D_k^{2/3} \\
	t'_k(D_k)=\alpha t_k= \alpha S_2(D_k)/2\epsilon
\end{array}\right.,
\end{equation}
%
where the only parameters in the model are $C$ (which is directly related to the Kolmogorov constant, and which we shall refer to as the Kolmogorov constant as well) and the persistence parameter $\alpha$.

For 3D-turbulence $C=C^{3D}=\frac{11}{3}C_2^{3D}$, with $C_2^{3D}\simeq 2.1$ the 3D-Kolmogorov constant (hence $C^{3D}\simeq7.7$). The model can also be applied to the inverse cascade regime of scales of 2D-turbulence, for which $S_2$ also obeys K41 scalings, with $C=C^{2D}=\frac{8}{3}C_2^{2D}$, where the 2D-Kolmogorov constant is $C_2^{2D}\simeq0.17 C_K^{2D}\simeq13.2$~(\cite{bib:lindborg1999_JFM}), with $C_K^{2D}\sim6$~(\cite{bib:boffetta2012_ARFM}) the spectral 2D-Kolmogorov constant (hence $C^{2D}\simeq35.3$).

The persistent parameter is the only adjustable parameter of the model. Based on numerical observations by~\cite{bib:bitane2012_PRE}, empirically suggesting $\alpha=1$, we shall also briefly discuss in the sequel this particular case (where $t'_k=t_k=S_2(D_k)/2\epsilon$) as a situation ``free of adjustable parameter'', although as pointed previously $\alpha=1$ is unlikely to be a relevant choice as it is incompatible with the ballisitic approximation. Note that in the iterative scheme similarly implemented by~\cite{bib:thalabard2014_JFM}, $\alpha=1$ is assumed (for each individual particle pair).

The next paragraphs show that the present scale dependent iterative scheme~(\ref{eq:cascade}) builds by itself a long term super-diffusive $t^3$ regime ``\`a la Richardson'' and gives a relevant description of the whole dispersion process, both for short term and long term regimes, as well as for the transition between one and the other. 

\subsection{From short-term ballisitic steps to long-term Richardson dispersion}\label{sec:richardson}
Substituting the explicit expressions for $S_2(D_k)$ and $t'_k(D_k)$ in (\ref{eq:cascade}) into the iteration equation for $D_k^2$, leads to a simple geometrical progression (and hence to an exponential growth of separation with the iteration number) both for the mean square separation $D_k^2$ and the ballisitic time scale $t'_k$:
%
\begin{subnumcases}{}
D_k^2={\cal A}^kD_0^2,\label{eq:Dk}\\
t'_k={\cal A}^{k/3}t'_0.\label{eq:tk}
\end{subnumcases}
with 
\begin{equation}\label{eq:A}
{\cal A}=1+\frac{\alpha^2C^3}{4}
\end{equation} 
and 
\begin{equation}
t'_0=\frac{\alpha C}{2}\epsilon^{-1/3}D_0^{2/3}.
\end{equation} 

From there, it is trivial algebra to relate the mean square separation at the iteration $k$ to the total iteration time $T_k=\sum_{j=0}^{k-1}t'_j$ (with $T_0=0$) :
\begin{equation}\label{eq:normDk}
\left(\frac{D_k^2}{D_0^2}\right)^{1/3}=1+({\cal A}^{1/3}-1)\frac{T_k}{t'_0},
\end{equation}

which can be equivalently written as 

\begin{equation}\label{eq:Dk2}
D_k^2=g\epsilon\left[T_k + \left(\frac{D_0^2}{g\epsilon}\right)^{1/3} \right]^3
\end{equation}

with 

\begin{equation}\label{eq:g}
g=\left[2\frac{{\cal A}^{1/3}-1}{\alpha C}\right]^3=\left[2\frac{(1+\frac{\alpha^2C^3}{4})^{1/3}-1}{\alpha C}\right]^3
\end{equation}

Before confronting these results to experimental and numerical data, several points are worth being briefly discussed :

\begin{itemize}
\item We can note that relations~(\ref{eq:Dk}) and~(\ref{eq:tk}) show that in the present model, particles dispersion eventually proceeds with a similar discrete scheme than in the GV04 model by~\cite{bib:goto2004_NJP} (although the GV04 model was originally derived only for 2D-turbulence, with a phenomenology related to separation bursts by specific hyperbolic points in the flow). It would be interesting in future studies to push further the comparison between both models. For instance, the relation ${\cal A}=1+\frac{\alpha^2C^3}{4}$ in the present model, suggests that the two parameters in the GV04 model can actually be related to each other, via the Kolmogorov constant. On the other hand, the GV04 phenomenology may help giving a physical interpretation, at least for the 2D-turbulence case, (in terms of the density of stagnation points for instance) for the persistence parameter $\alpha$ introduced here on mainly kinematic considerations (to warrant the ballisitic approximation validity). Similarly, the question of a possible phenomenological connection between both approaches in 3D-turbulence may also be addressed.


\item Expression~(\ref{eq:Dk2}) shows that at short term, dispersion may still exhibit an apparent Richardson-like cubic regime $D_k^2=g\epsilon(T_k-T_{origin})^3$, as long as a negative virtual time origin is introduced : $T_{origin}=-(D_0^2/g\epsilon)^{1/3}=-2t_0/Cg^{1/3}$, which depends on the initial separation $D_0$. This justifies and validates the necessity of introducing such a negative time shift in the experiments by~\cite{bib:ott2000_JFM}, to devise a cubic regime from the available short time measurements. This also suggests that the experimental data by Ott \& Mann can be fitted with only one free parameter (namely the Richardson constant $g$) as the time shift $T_{origin}$ only depends on $g$ and the known initial separation and energy dissipation rate. It would be interesting to revisit this experimental data with this vision.

\item Expression~(\ref{eq:Dk2}) has an asymptotic long-term behavior $D_k^2=g\epsilon T_k^3$, 
showing that the elementary scale dependent ballisitic steps, trivially build a cubic regime, independent on initial separation, \emph{\`a la } Richardson. Besides, eq.~(\ref{eq:g}) relates the Richardson constant $g$ to the two parameters of the ballisitic phenomenology (namely the Kolmogorov constant $C$ and the persistence parameter $\alpha$).

\end{itemize}

\section{Results of the model and comparison with existing numerical and experimental data}

First, the practical implementation of the model, requires the 2 parameters $C$ and $\alpha$ to be prescribed. 

\subsection{Determination of relevant values of model parameters}
\begin{figure}
\center
\begin{minipage}[c]{.46\linewidth}
\center
\includegraphics[height=5cm]{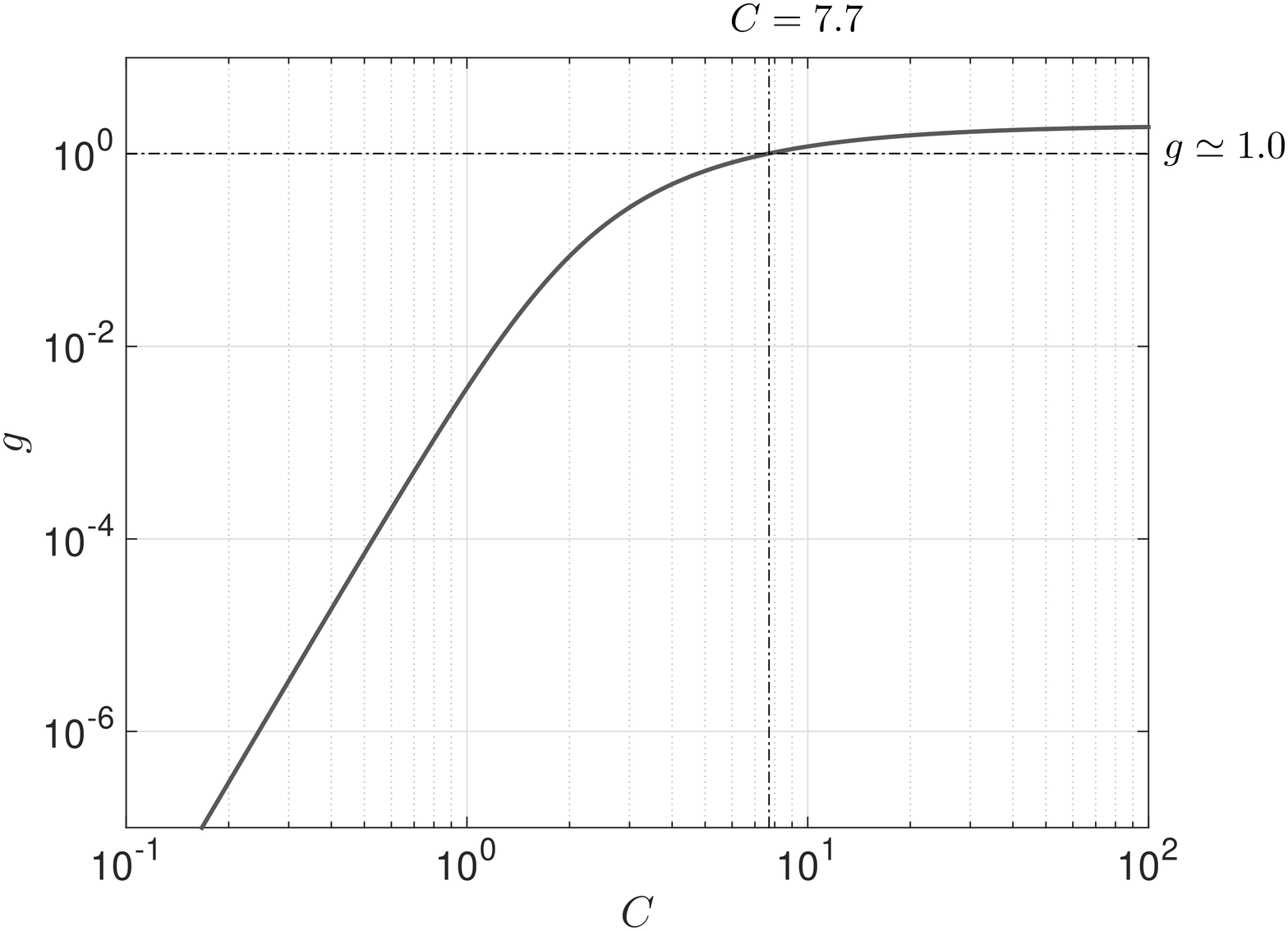}\\
(a)
\end{minipage}
\hfill
\begin{minipage}[c]{.46\linewidth}
\center
\includegraphics[height=5cm]{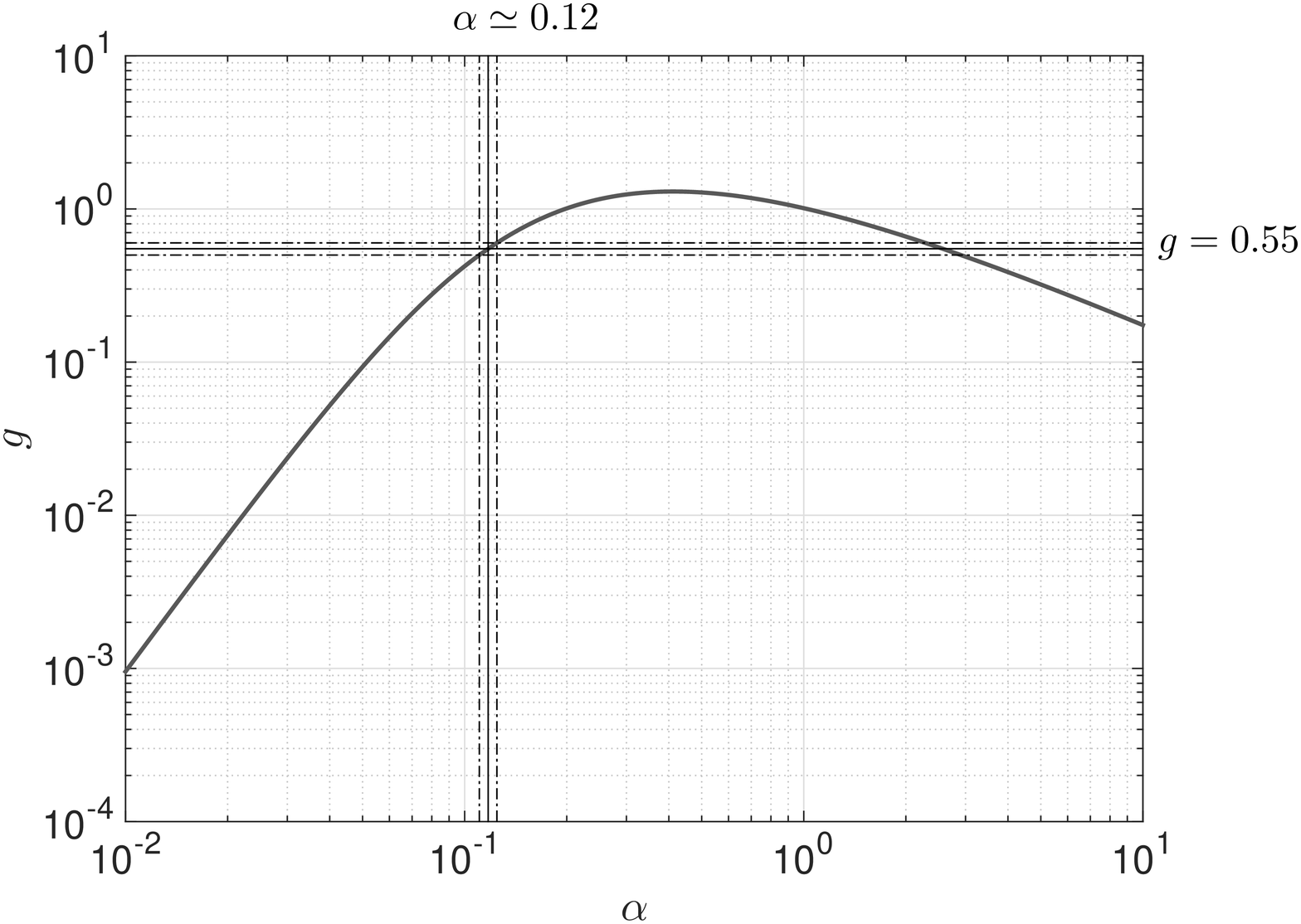}\\
(b)
\end{minipage}

\caption{(a) Dependency of the Richardson constant $g$ on the Kolmogorov constant $C$ in the ballisitic mode when the persistence parameter is fixed to $\alpha=1$. (b) Prediction of the Richardson constant $g$ on the persistence parameter $\alpha$, when the Kolmogorov constant is fixed to $C=C^{3D}=7.7$.}\label{fig:g_C}
\end{figure}

We shall focus in the sequel on the case of 3D-turbulence, hence taking for the parameter $C$ the value $C^{3D}=\frac{11}{3}C_2^{3D}\simeq7.7$. Later, in section~\ref{sec:asymmetry}, we shall also briefly discuss some aspects of the inverse cascade regime of 2D-turbulence, in which case $C^{2D}=\frac{8}{3}C_2^{2D}\simeq35.3$ will be used for $C$.

If we forget for a moment about the persistence parameter (hence considering the typical ballisitic time scale as being simply $t_k=S_2(D_k)/2\epsilon$, assuming $\alpha=1$, as empirically suggested by Bitane~\emph{et al.} simulations and as done in Thalabard~\emph{et al.}), relation~(\ref{eq:g}) then directly connects the Richardson constant $g$ to the Kolmogorov constant only. Figure~\ref{fig:g_C}a shows the predicted evolution of $g$ as a function of $C$ for the case $\alpha=1$. Interestingly this figure shows that the predicted value for $g$ when $C=C^{3D}\simeq 7.7$ is then $g^{3D}\simeq1.0$. It is appealing that, although this value is slightly larger than the well accepted value $g^{3D}\simeq 0.5-0.6$, it is still a reasonable estimate considering the simplicity of the model and the lack of adjustable parameter. However, figure~\ref{fig:g_C}a also shows that for $\alpha=1$, $g$ tends asymptotically to the limit $g_\infty=2$ as the Kolmogorov constant increases. This clearly shows a physical limit of the case $\alpha=1$ (beyond the previous considerations requesting $\alpha < 1$ for the ballistic approximation to hold), as numerical simulations of pair dispersion in the inverse cascade regime of 2D-turbulence report values of $g^{2D}$ larger than 2 (\cite{bib:boffetta2002_PoF} report for instance a value $g^{2D}\simeq3.8$ while \cite{bib:faber2009_PoF} report $g^{2D}\simeq6.9$). This observation, raised by one of the anonymous referees of the first version of this article, also supports the necessity of considering the persistence parameter $\alpha$. 

To determine the optimal value of the persistence parameter $\alpha$ for the case of 3D-turbulence, we therefore use relation~(\ref{eq:g}) with the value for the Kolmogorov constant prescribed to $C=C^{3D}\simeq7.7$ and we seek for the value of $\alpha$ for which the well accepted value for the Richardson constant, $g=g^{3D}=0.55\pm0.05$, is retrieved.  Figure~\ref{fig:g_C}b shows the dependency of $g$ predicted by the iterative ballistic phenomenology as a function of the persistence parameter $\alpha$ for $C=C^{3D}\simeq7.7$. We see that $g\simeq0.55\pm0.05$ for $\alpha\simeq0.118\pm0.007$ (a second solution exists for $\alpha\simeq2.57 > 1$ which is out of the range of validity of the ballistic approximation, requiring $\alpha < 1$ and which will therefore not be considered). The value $\alpha = 0.12$ will therefore be used in the sequel for the persistence parameter in 3D-turbulence. It can also be noted in figure~\ref{fig:g_C}b that the evolution of $g$ with $\alpha$ has a mild maximum around $\alpha\simeq 0.4$, so that the dependency of $g$ on $\alpha$ is relatively weak in the range $0.1<\alpha<1$, where the Richardson spans the relatively narrow range $0.4<g<1.3$.

\begin{figure}
\center
\includegraphics[height=7cm]{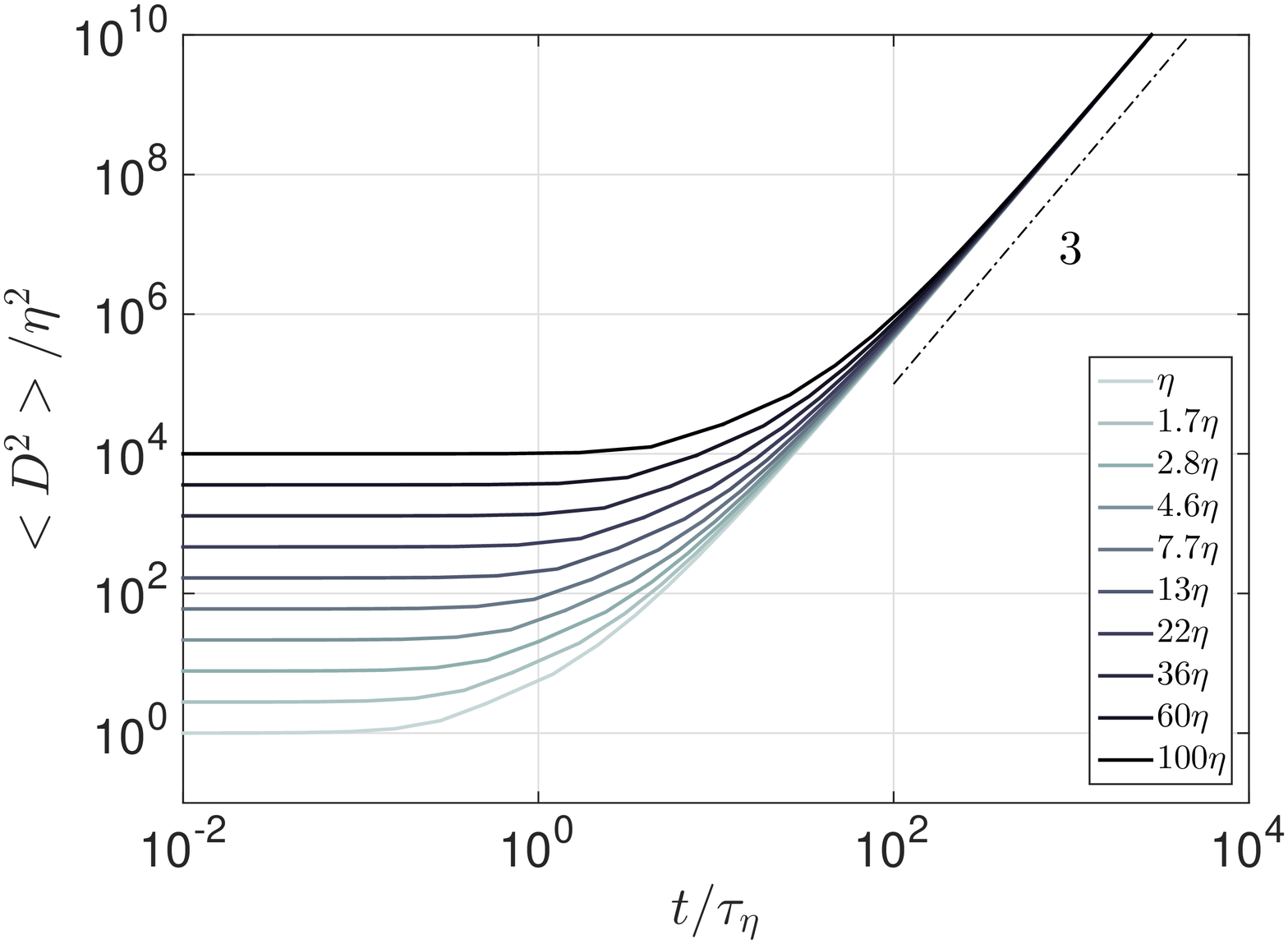}
\begin{picture}(0,0)
\put(-240,110){\includegraphics[height=2.4cm]{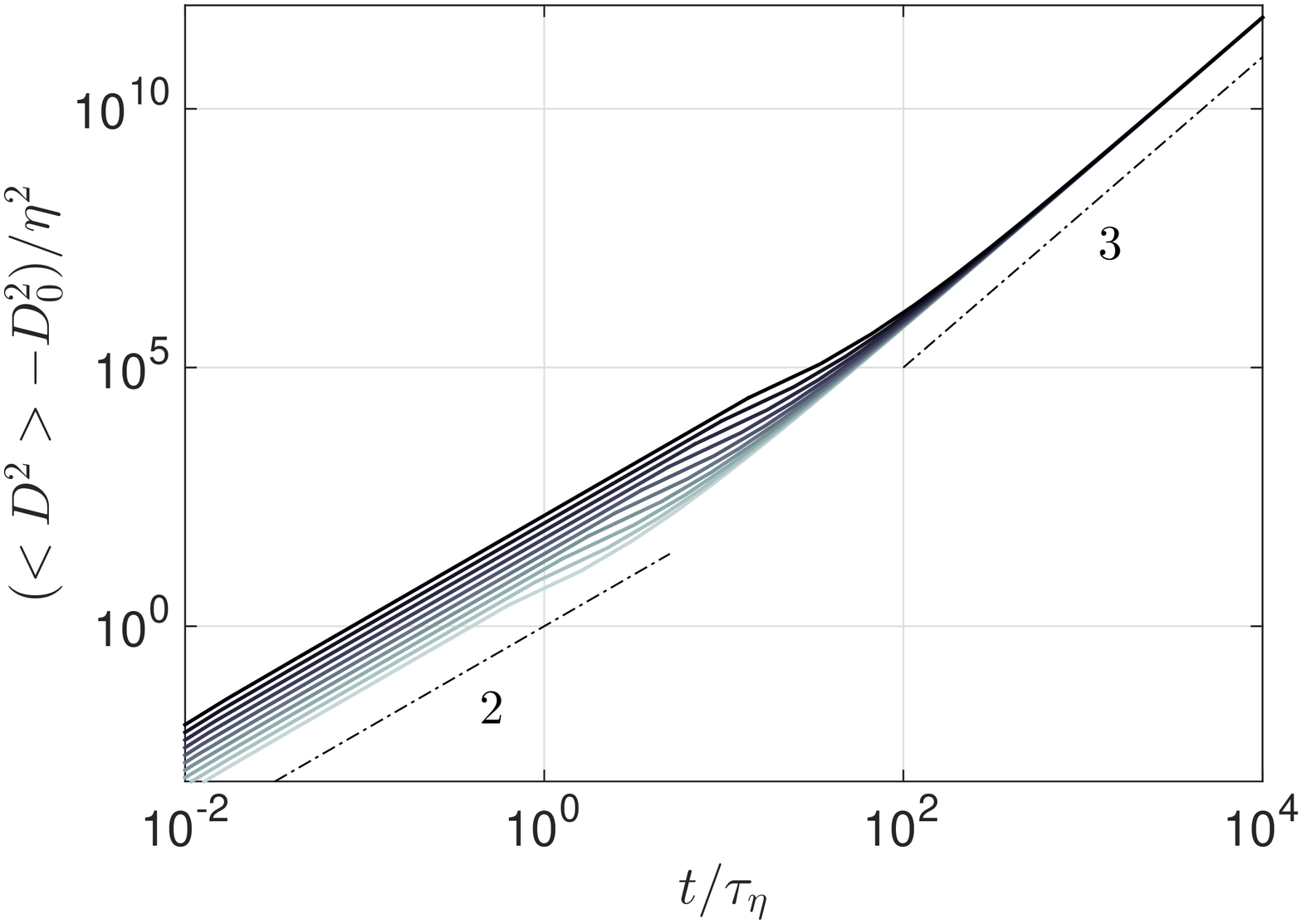}}
\end{picture}
\caption{Growth of the mean square separation between particles predicted as the iterative scale dependent ballistic process (defined by eq.~\ref{eq:cascade}) propagates starting from several initial separations $D_0$ in the inertial range of turbulence. The mean square separation $<D^2>$ is normalized by the square of the dissipative scale $\eta^2$ and time is normalized by the dissipative time scale $\tau_\eta$. Turbulent energy dissipation rate and dissipative scales have been choosen to match experiments in~[\cite{bib:bourgoin2006_Science}], although as shown in figure~\ref{fig:D2T2}, the results are independent of this particular choice when the data is properly normalized. It can be seen in this plot that at long times, a dispersive regime ``\`a la Richardson'', cubic in time and independent of initial separation, is naturally built by the iterative scale dependent ballistic process.}\label{fig:R2_t}
\end{figure}

\subsection{Practical implementation of the model}\label{sec:practical}

Let me now first illustrate the proposed ballistic cascade mechanism, in 3D turbulence, using the values $C^{3D}=7.7$ and $\alpha=0.12$ for parameters of the model, and with realistic numbers for the energy dissipation rate and initial separations compared to existing experiments and simulations. Concerning the energy dissipation, I use the value corresponding to the experiment by\cite{bib:bourgoin2006_Science} at $R_\lambda=815$: $\epsilon\simeq6.25$~m$^2$s$^{-3}$ (giving a dissipation scale $\eta\simeq 23\mu$m for a flow of water and a dissipation time scale $\tau_\eta\simeq 0.4$~ms). Several initial separations $D_0$, spanning the range $[\eta ; 100\eta]$ will be considered. 

Figure~\ref{fig:R2_t} shows the mean square separation obtained then from the iterative ballisitc scheme given by eqs.~(\ref{eq:cascade}). It can be seen that in the long term the iterative ballistic process eventually leads to a cubic super-diffusive regime ``\`a la Richardson'', independent on the initial separation. The inset shows the same data but with the initial mean square separation $D_0^2$ substracted in order to better emphasize the initial ballistic regime at short times and the transition toward the Richardson regime. Note that time $t$ in these plots corresponds to the total discrete iteration time $T_k$ previously defined. 

In figure~\ref{fig:R2_t} time is non-dimensionalized using the dissipative time scale $\tau_\eta$ while mean square separations are non-dimensionalized using the square of dissipative scale $\eta^2$. However, the natural scales in the iterative ballistic process proposed here are $t_0$ (or $t'_0=\alpha t_0$) for the time scale and $D_0$ for the spatial scale. Relation~(\ref{eq:normDk}) shows indeed that with such non-dimensionalized separations and time, the separation process only depends on the parameters $C$ and $\alpha$ of the model, via the constant $\cal A$ (eq.~\ref{eq:A}). For the sake of comparison with numerical results I use the same non-dimensionalization as \cite{bib:bitane2012_PRE}, based on $t_0^2S_2(D_0)=\frac{1}{2}C^3D_0^2$ (rather than just $D_0^2$, without loss of generality) for the mean square separation and on $t_0$ for the time. With such a choice, relation~(\ref{eq:Dk2}) for the evolution of the mean square separation simply becomes :
\begin{equation}\label{eq:Xk2}
X_k^2 = \frac{g}{2}\left[\tau_k - \frac{2}{C g^{1/3}} \right]^3
\end{equation}
with $X_k^2=D_k^2/t_0^2 S_2(D_0)$ and $\tau_k=T_k/t_0$.

\begin{figure}
\center
\includegraphics[height=7cm]{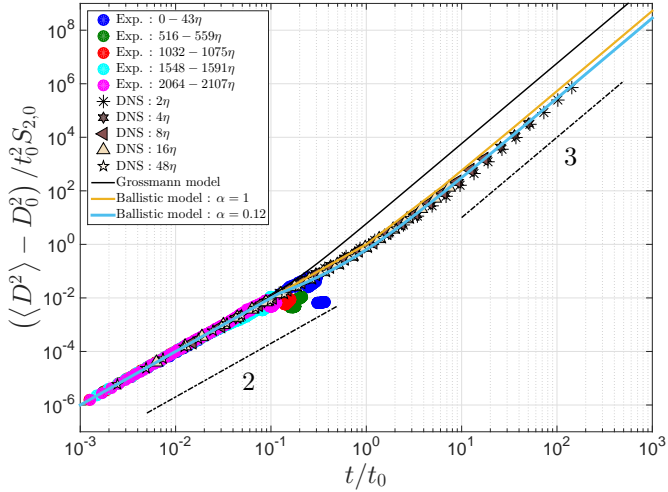}
\caption{Growth of the mean square separation (with the initial separaion substracted) normalized by $t_0S_2(D_0)$ as a function of time normalized by $t_0$. The light blue solid line correspond to the exact same data as in figure~\ref{fig:R2_t} from the ballistic model (with $C=7.7$ and $\alpha=0.12$), which with the present normalization collapses onto a single curve, according to relation~\ref{eq:Xk2}. The yellow line corresponds to the prediction of the ballistic model with $C=7.7$ and $\alpha=1$. Colored circles correspond to experimental results for the mean square pair separaion in~[\cite{bib:bourgoin2006_Science}] for different intial separations. Other symbols correspond to the mean square separation for different intial separations from the DNS by Bitane \emph{et al.}~[\cite{bib:bitane2012_PRE}]. Finally the solid black line shows the prediction from Grossmann and Procaccia model~[\cite{bib:grossmann1984_PRA,bib:grossmann1990}]. Two important aspects to be noted are : (i) the limited temporal range of experimental data, which does not allow to probe the transition between the initial ballisitc regime and the late Richardson cubic regime and (ii) the remarkable agreement between the ballisitic model and the numerical simulations.}\label{fig:D2T2}
\end{figure}

The light blue line in figure~\ref{fig:D2T2} represents the same data as in figure~\ref{fig:R2_t}, but with this new non-dimensionalization, which now collapses the mean square separations for all initial separations into a single curve. I have also reported on the same figure the mean square separations obtained for various initial separations in the simulations by \cite{bib:bitane2012_PRE}. The agreement of the iterative ballisitic phenomenology compared to the numerics is almost perfect. The first interesting observation concerns the collapse of the numerical data with this normalization (as already noted by~\cite{bib:bitane2012_PRE}) which is perfectly reproduced by the present phenomenology. Besides, not only the global trend of the mean square separation evolution is very well descibed by the model (both for short and long term regimes), but some subtler details are also well captured. For instance, as in the simulation, the transition between the early Batchelor regime and the Richardson regime is robustly found to occur around $t=t_0$, even if the duration of the initial ballistic iteration is $t'_0=\alpha t_0$ (hence significantly smaller than $t_0$ in the present case). A mild slowing down of the ballistic separation for $t\lesssim t_0$ (prior to the transition towards the cubic regime), present in the simulations is also captured by the iterative ballistic model.

For comparison, figure~\ref{fig:D2T2} also shows the prediction of the ballistic model, with unity persistence parameter $\alpha=1$, hence taking $t'_k=t_k=S_2(D_k)/2\epsilon$ as the characteristic time of the fundamental ballistic process (as originally proposed by~\cite{bib:bitane2012_PRE} and in the line of the implementation of the ballistic phenomenology by Thalabard~\emph{et al.}). It is interesting to note, that although the long term separation is then slightly over-estimated (as a result of the Richardson constant being $g=1.0$ in the model with $\alpha=1$ as previously discussed, instead of $g\simeq 0.55$), the global picture for the mean-square separation is still relatively well captured, with no other additional adjustable parameter in this case than the Kolmogorov constant. Overall, the global picture of the temporal growth of the mean-square separation in the present ballistic phenomenology is not very sensitive to the specific value of the persistence parameter (as long as we remain in the range $0.1 < \alpha < 1.0$) : when the persistence parameter is varied in the range $0.1 < \alpha < 1.0 $ the Batchelor to Richardson transition is always observed to occur for times around $t_0$, with a Richardson constant restricted to the range $0.4<g< 1.3$ (see figure~\ref{fig:g_C}b), reasonable compared to the well accepted value of 0.55. The choice $\alpha\simeq 0.12$, remains however optimal in terms of quantitative comparison with available numerical data for the mean square separation.

Figure~\ref{fig:D2T2} also reports the prediction for the mean square separation of particle pairs from an altenative model of turbulent relative dispersion, proposed by Grossmann \& Procaccia in 1984~(\cite{bib:grossmann1984_PRA,bib:grossmann1990}). This model relies on a mean field approach of Navier-Stokes equation and a phenomenological closure assumption to predict the evolution of the mean square sepration of pairs of particles (with initial separation within inertial scales of turbulence) as follows  :
\begin{equation}\label{eq:grossmann}
	\left< \left(\vec{D}-\vec{D}_0 \right)^2\right>=\left(D_0^{4/3}+\frac{2}{3}C\epsilon^{2/3}t^2\right)^{3/2}-D_0^2
\end{equation}
whose long time approximation asymptotically reaches a cubic Richardson regime $<(\vec{D}-\vec{D}_0 )^2> = g'\epsilon t^3$ with $g'=\left(\frac{2}{3}C\right)^{3/2}$. Interestingly this model also proposes a direct connection between the Richardson and the Kolmogorov constant $C$. However, although the prediction given by eq.~(\ref{eq:grossmann}) and shown in figure~\ref{fig:D2T2} captures correctly the initial ballistic separation, it significantly anticipates the transition towards the Richardson regime which occurs much earlier than in the DNS or the iterative ballistic phenomenology proposed, what results in a significant over-estimation of the actual mean-square separation at long times. This sooner transition is related to the over-estimation of the Richardson constant in Grossmann \& Procaccia's model : $g'=\left(\frac{2}{3}C\right)^{3/2}\simeq 12$ (with $C=7.7$), what is much larger than the well accepted value $g\simeq 0.55$.

Finally, I have also superimposed in figure~\ref{fig:D2T2} the mean square separation of pairs measured in high resolution particle tracking experiments by~\cite{bib:bourgoin2006_Science}. In those experiments, only the Batchelor ballisitc regime was reported, while no hint of Richardson regime was detected. Figure~\ref{fig:D2T2} emphasizes a possible reason for the failure in experiments to observe the Richardson regime : the longest experimental tracks did not exceed a few tenth of $t_0$ while the separation needs to be tracked for at least a few $t_0$ to reasonably detect the transition toward the cubic regime. A simple possible strategy to improve the chances to observe the cubic regime in experiments would simply consist in better controlling the injection of particle pairs in order to achieve sufficiently small initial separations, hence reducing the time $t_0$ required for the transition to occur within experimentally accessible tracking time.

\section{Temporal asymmetry of turbulent dispersion}\label{sec:asymmetry}
As pointed in the introduction, a noticeable feature of turbulent dispersion is the temporal asymmetry. This means that backward and forward dispersion operate at different rates, as pointed by~\cite{bib:sawford2005_PoF}. In 3D turbulence, backward dispersion has for instance been shown numerically and experimentally to operate twice as fast as forward dispersion (see \cite{bib:berg2006_PRE,bib:bragg2014_arXiv}). In the inverse cascade regime of 2D turbulence, the opposite trend was reported by~\cite{bib:faber2009_PoF}.

In its present formulation, the iterative ballistic phenomenology is completely time-reversible as the elementary ballistic process in eq.~(\ref{eq:disp1a}) at each scale is quadratic in time and hence fully reversible under the transformation $t\rightarrow -t$. I propose in this section a simple extension of the iterative ballistic phenomenology to address the question of time assymmetry.

\subsection{A ballistic cascade phenomenology}
Time irreversibility can be simply introduced by pushing up to third order the Taylor expansion leading to the elementary ballistic process~(\ref{eq:disp1a}), as already presented in eq.~(\ref{eq:3rdOrder}). When pushed to third order, the iterative scheme~(\ref{eq:cascade}), giving the growth of pair separation at iteration between the $k^{th}$ and the $(k+1)^{th}$ iteration then becomes :

\begin{equation}\label{eq:O3}
D_{k+1}^2=D_k^2+S_2(D_k) t^{\prime 2}_k+S_{au}(D_k) t^{\prime 3}_k,
\end{equation}
where $S_{au}(r)=\left<\delta_{\vec{r}}\vec{a}\cdot\delta_{\vec{r}}\vec{u}\right>$ is the crossed velocity-acceleration strcture function.

The relevance of considering the cubic term in the iterative process is also supported by the fact that in the practical implementation of the iterative ballisitic phenomenology discussed in the previous section, the optimal value for the persistence parameter $\alpha$ was found to be of the order of 0.1, while as discussed in section~\ref{sec:ballistic} neglecting completely the third order term would require $\alpha \ll 1$. With $\alpha\simeq 0.1$, the cubic term can still be expected to contribute of the order of 10\% to the quadratic separation at each iteration step. It is therefore reasonable to consider a possible corrective contribution of third order effects in the ballistic phenomenlogy.

From a physical point of view, the third order term has a clear energetic interpretation. As discussed in section~\ref{sec:ballistic}, the crossed velocity-acceleration structure function $\left<\delta_{\vec{r}}\vec{a}\cdot\delta_{\vec{r}}\vec{u}\right>$ in 3D turbulence is predicted to be $-2\epsilon$ (at all inertial scales), and hence solely related to the energy dissipation rate $\epsilon$. If we were discussing single particle issues (instead of particle pairs) this would be simply understood as a dissipative correction of the average ballistic motion of the particle, as $\left<\vec{a}\cdot\vec{u}\right>=\frac{1}{2}\left<\frac{\textrm{d}u^2}{\textrm{d}t}\right>$ is indeed the average dissipation of particle kinetic energy along its Lagrangian path. When the relative motion of particles in a pair is considered, the physical meaning of the crossed velocity-acceleration structure function $S_{au}=\left<\delta_{\vec{r}}\vec{a}\cdot\delta_{\vec{r}}\vec{u}\right>$ is however somehow subtler. $S_{au}$ can indeed be analytically related to the third order velocity structure function (and hence to the energy cascade accross scales) directly from Navier-Stokes equation, such that under local stationarity and homogeneity assumptions~(see for instance \cite{bib:risoeReport,bib:hill2006_JoT}) :
\begin{equation}
	 2\left<\delta_{\vec{r}}\vec{a}\cdot\delta_{\vec{r}}\vec{u}\right> = \vec{\nabla} \cdot \left< \delta_{\vec{r}} \vec{u} \delta_{\vec{r}} \vec{u} \cdot \delta_{\vec{r}} \vec{u} \right>.
\end{equation}

In 3D turbulence, $\vec{\nabla} \cdot \left< \delta_{\vec{r}} \vec{u} \delta_{\vec{r}} \vec{u} \cdot \delta_{\vec{r}} \vec{u} \right>=-4\epsilon$ (what is an exact relation and an alternative version of the Karman-Howarth-Monin relation under local homogeneity and isotropy assumptions~\cite{bib:frisch}), what retrieves the relation $S_{au}^{3D}=-2\epsilon$ previously mentioned for the case of 3D turbulence. The negative sign in these relations reflects the fact that in 3D turbulence the energy cascade at inertail scales is a \emph{direct cascade} (energy flows from large to small scales).

Conversely, in the inverse cascade of 2D turbulence $\vec{\nabla} \cdot \left< \delta_{\vec{r}} \vec{u} \delta_{\vec{r}} \vec{u} \cdot \delta_{\vec{r}} \vec{u} \right>=+4\epsilon$~(\cite{bib:lindborg1999_JFM}), what leads then to $S_{au}^{2D}=+2\epsilon$. The positive sign refering now to the \emph{inverse} nature of the energy cadcade.

The crossed velocity-acceleration structure function therefore carries the signature of the energy flux accross scales. In the present context of pair dispersion, the cubic term in eq.~(\ref{eq:O3}) should therefore be seen as the Lagrangian signature of the energy flux due to the relative motion (relative velocity and relative acceleration) of particles separating from a given scale to larger scales. The negative sign in the relation $S_{au}^{3D}=-2\epsilon$ in 3D can be interpreted as the fact that in the forward disperion process, as particles separate (from small to large scales), they climb the energy cascade ``upstream'', \emph{against} the energy flux (which flows from large to small scales in the direct 3D cascade), while in the backward case separation pair separation climbs the cascade downstream, \emph{with} the energy flux. The scenario is reversed in the inverse cascade of 2D turbulence.

The time asymmetry introduced by the third order correction to the ballistic process in the extended iterative phenomenology therefore accounts, in the Lagrangian framework of pair dispersion, for the scale asymmetry of energy flux in the turbulent cascade. We shall therefore refer to this extension as the~\emph{ballistic cascade phenomenology}.

We can note, that when the relation $S_{au}^{3D}=-2\epsilon$ (for 3D turbulence) is reported in the elementary short term separation process given by eq.~(\ref{eq:3rdOrder}) (or by eq.~\ref{eq:O3}), we find that for short times, the short term separation is naturally dominated by the ballistic (quadratic) contribution, with a third order temporal asymmetry such that the difference between short term forward and backward separation is $<D^2>(-t)-<D^2>(t)=4\epsilon t^3$ (in 3D turbulence). This corresponds to the short term cubic in time asymmetry of relative dispersion recently investigated by~\cite{bib:jucha2014_PRL}. We will show here (in next sub-section) that the iterative propagation of this short term asymmetry also builds the long-term asymetry, in quantitative agreement with previous studies for the Richardson regime asymmetry. Such a connection between the short term and long term asymmetry still remained to be established, as pointed by~\cite{bib:jucha2014_PRL}.

\subsection{Explicit formulation of the ballistic cascade phenomenology}

The extended phenomenology can be explicitly implemented as follows 
\begin{equation}\label{eq:O3total}
D_{k+1}^2=D_k^2+S_2(D_k) t^{\prime 2}_k+S_{au} t^{\prime 3}_k  \;\; \textrm{ with }\;\; \left\{
\begin{array}{l}
	S_2(D_k)=C\epsilon^{2/3}D_k^{2/3} \\
	S_{au} = C_{au} 2 \epsilon\\
	t'_k(D_k)=\alpha t_k= \alpha S_2(D_k)/2\epsilon
\end{array}\right.,
\end{equation}
where the asymmetry coefficient $C_{au}=\pm1$ depending on the inverse or direct nature of the energy cascade ($C_{au}^{3D}=-1$ for the 3D cascade in 3D turbulence and $C_{au}^{2D}=+1$ for the inverse cascade in 2D turbulence). I also recall the values for the Kolmogorov constant (previously discussed in section~\ref{sec:practical}) : $C^{3D}=7.7$ and $C^{2D}=35.3$.

Time asymmetry between forward and backward dispersion can then be simply considered by changing $t'_k\rightarrow -t'_k$ (what only affects the third order term in~(\ref{eq:O3total})) or equivalently by reversing the sign of $C_{au}$ when propagating the iterative scheme. The following table specifies the values to be used for the asymmetry parameter $C_{au}$ for the different cases :\\
\begin{center}
$C_{au}$\\
\begin{tabular}{ccc}
\hline
	&	3D 	&	2D \\
	&	(direct cascade)	&	(inverse cascade)\\
	\hline
	\hline
 forward 		& 	-1	&	+1\\
 backward		&	+1	&	-1\\
 \hline
\end{tabular}
\end{center}

\subsection{From short term asymmetric ballistic dispersion to long term asymmetric Richardson dispersion}
As for the case of the purely ballistic model discussed in section~(\ref{sec:richardson}), the new iterative scheme can be explicitly solved by substituting the expressions for $S_2$, $S_{au} $ and $t'_k$ into the iterative relation in~(\ref{eq:O3}). Since $S_{au}$ does not depend on the separation $D_k$, the solution is straightforward, and leads to a geometric progression for $D_k^2$ and $t'_k$, analog to the previous relations~(\ref{eq:Dk})~and~(\ref{eq:tk}) :

\begin{subnumcases}{}
D_k^2={\cal A}^{\prime k}D_0^2,\label{eq:Dk}\\
t'_k={\cal A}^{\prime k/3}t'_0.\label{eq:tk}
\end{subnumcases}
with 
\begin{equation}\label{eq:A'}
{\cal A}'=1+\frac{\alpha^2C^3}{4}\left(1+C_{au}\alpha\right).
\end{equation}
Note that setting $C_{au}=0$ (hence neglecting the third order term) naturally retrieves ${\cal A}'={\cal A}=1+\frac{\alpha^2C^3}{4}$ (as given by eq.~\ref{eq:A} for the purely ballistic model).

The temporal evolution of the mean square separation can therefore be written with an analogous formulation as eqs.~(\ref{eq:Dk2})~\&~(\ref{eq:g}) 
\begin{equation}\label{eq:Dk2b}
D_k^2=g\epsilon\left[T_k + \left(\frac{D_0^2}{g\epsilon}\right)^{1/3} \right]^3
\end{equation}

with 

\begin{equation}\label{eq:gO3}
g=\left[2\frac{{\cal A}^{\prime1/3}-1}{\alpha C}\right]^3=\left[2\frac{(1+\frac{\alpha^2C^3}{4}\left(1+C_{au}\alpha\right))^{1/3}-1}{\alpha C}\right]^3
\end{equation}

As a result, the third order corrected model behaves exactly as the purely ballistic model, except for the relation between the Richardson constant and the model parameters which now includes an $\alpha^3$ corrective term in the expression of the constant ${\cal A}'$ (eq.~\ref{eq:A'}), associated to the asymmetry coefficient $C_{au}$.

\begin{figure}
\center
\includegraphics[height=6cm]{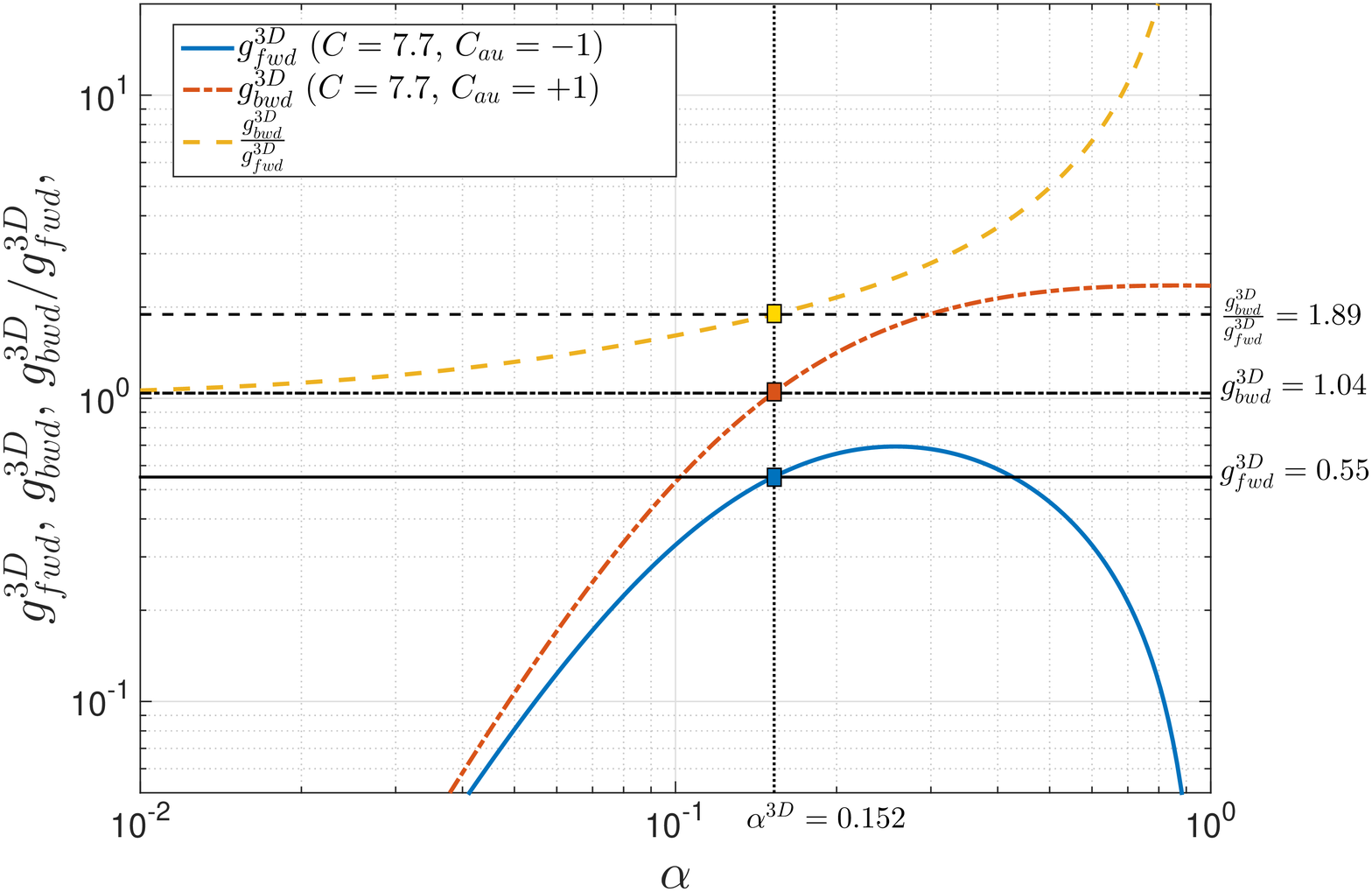}\\
\caption{Dependency of the Richardson constant predicted by the ballistic cascade model (with third order term correction, eq.~\ref{eq:O3total}) for the forward and backward dispersion problem in direct cascade 3D turbulence, as a function of the persistence parameter. The plot shows the forward Richardson constant $g^{3D}_{fwd}$ (plain blue line), the backward Richardson constant $g^{3D}_{bwd}$ (red dot-dashed line) and the ratio $g^{3D}_{bwd}/g^{3D}_{fwd}$ (yellow dashed line). The vertical dot line emphasizes the optimal value for the persistence parameter, determined as the smallest value of this parameter for which the forward Richardson constant predicted by the model matches the  values in the litterature ($g^{3D}_{fwd}\simeq0.55$ ). The horizontal lines emphasize the corresponding values for the forward and backward Richardson constants, as well as for the ratio $g^{3D}_{bwd}/g^{3D}_{fwd}$ (these values are reported on the right side of the plots).}
\label{fig:gbwd_vs_gfwd}
\end{figure}

\subsection{Forward vs Backward dispersion}
I consider here the practical implementation of the ballistic cascade phenomenology, exploring the consequences in terms of temporal asymmetry of the dispersion process in the direct cascade of 3D turbulence and the inverse cascade regime of 2D turbulence.
\subsubsection{The case of direct cascade in 3D turbulence}
Figure~\ref{fig:gbwd_vs_gfwd}a represents the dependency on the persistence parameter $\alpha$ of the Richardson constant as predicted by the ballistic cascade phenomenology given by relation~(\ref{eq:gO3}) for the case of 3D turbulence ($C=7.7$) in the forward ($C_{au}=-1$) and backward ($C_{au}=+1$) situations ($g^{3D}_{fwd}$ and $g^{3D}_{bwd}$ are respectively the 3D-forward and 3D-backward Richardson constants). The figure also shows the ratio $g^{3D}_{bwd}/g^{3D}_{fwd}$ as a function of $\alpha$. This ratio is found to be always larger than unity, what shows that for a given value of the persistence parameter, the backward dispersion in 3D turbulence always operates faster than forward dispersion, in qualitative agreement with experiments and simulations by~\cite{bib:berg2006_PRE}. In the context of the present phenomenology, the faster separation in the backward case compared to the forward case is directly related to the positiveness and negativeness of $C_{au}$ in each respective situation. The positive value of $C_{au}$ in the backward case, results in an enhancement of the separation due to the positive third order corrective term in~\ref{eq:O3total} which accelerates \emph{in fine} the long term cubic separation compared to the case of the purely ballistic iterative phenomenology. On the contrary, the negative value of $C_{au}$ for the forward case results in the reduction of the separation due to the negative third order corrective term, which decelerates the long term cubic separation compared to the purely ballistic iterative phenomenology. 

The comparison with experiments and simulations by~\cite{bib:berg2006_PRE} can be pushed to a quantitative level by considering the optimal value of the persistence parameter, for which $g^{3D}_{fwd}=0.55$ (the well accepted value for the forward dispersion problem). It can be seen in figure~\ref{fig:gbwd_vs_gfwd}a that two values of $\alpha$ satisfy this condition, one around 0.15, the other around 0.42. There is no \emph{a priori} obvious choice to select one or the other of these values, except that the validity of the short term Taylor expansion for the mean square separation (which is the starting point of the present phenomenology) is expected to be more robust for shortest times, what tends to prefer the smallest compatible value for the persistence parameter. We will therefore take $\alpha\simeq0.15$ as the optimal value for the persistence parameter, compatible with the well accepted value $g^{3D}_{fwd}=0.55$. Figure~\ref{fig:gbwd_vs_gfwd}a then shows that the corresponding value for the backward Richardson constant is $g^{3D}_{bwd}\simeq1.04$, leading to a ratio $g^{3D}_{bwd}/g^{3D}_{fwd}\simeq 1.9$, in good quantitative agreement with the experiments and simulations by~\cite{bib:berg2006_PRE} and with the more recent simulations by~\cite{bib:bragg2014_arXiv}, who all find a ratio $g^{3D}_{bwd}/g^{3D}_{fwd}\simeq2$.

\begin{figure}
\center
\includegraphics[height=6cm]{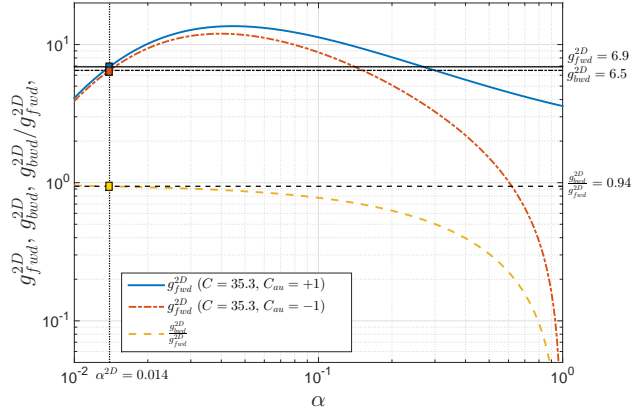}\\
\caption{Same as figure~\ref{fig:gbwd_vs_gfwd}, but for the case of 2D inverse cascade. The optimal value for the persistence parameter, is here determined as the smallest value for which the forward Richardson constant predicted by the model matches the value in~\cite{bib:faber2009_PoF} ($g^{2D}_{fwd}\simeq6.9$).}
\label{fig:gbwd_vs_gfwd_2D}
\end{figure}

\subsubsection{The case of inverse cascade in 2D turbulence}
Figure~\ref{fig:gbwd_vs_gfwd}b represents the dependency on the persistence parameter $\alpha$ of the Richardson constant as predicted by the ballistic cascade phenomenology for the case of 2D turbulence ($C=35.3$) in the forward ($C_{au}=+1$) and backward ($C_{au}=-1$) situations ($g^{2D}_{fwd}$ and $g^{2D}_{bwd}$ are respectively the 2D-forward and 2D-backward Richardson constants). The figure also shows the ratio $g^{2D}_{bwd}/g^{2D}_{fwd}$ as a function of $\alpha$. This ratio is found to be always smaller than unity, what shows that for a given value of the persistence parameter, the backward dispersion in the inverse cascade regime of 2D turbulence always operates slower than forward dispersion, in qualitative agreement with the numerical simulations by~\cite{bib:faber2009_PoF}. This trend is the opposite than in 3D turbulence, what is directly related to the fact that $C_{au}$ has opposite sign for the inverse and direct cascade situations.

For a more quantitative analysis of the 2D case, we proceed as previously by determining the optimal value of the persistence parameter $\alpha$ compatible with reported values for the forward Richardson constant $g^{2D}_{fwd}$. As discussed in the introduction, values for $g^{2D}_{fwd}$ in the litterature still span a broad range. Experiments by~\cite{bib:jullien1999_PRL} suggest $g^{2D}_{fwd}\simeq0.55$, while numerical simulations by~\cite{bib:boffetta2002_PoF} and by~\cite{bib:faber2009_PoF} report $g^{2D}_{fwd}\simeq3.8$ and $g^{2D}_{fwd}\simeq6.9$ respectively. For the sake of the present discussion, I will consider the value proposed by Faber \& Vassilicos ($g^{2D}_{fwd}\simeq6.9$) as this study also addresses explicitly and quantitatively the comparison between forward and backward dispersion in the inverse cascade regime of 2D turbulence. Figure~\ref{fig:gbwd_vs_gfwd}b shows that, as for the 3D case, two possible values of $\alpha$ are compatible with the value $g^{2D}_{fwd}\simeq6.9$, one around 0.014, the other around 0.27. As for the 3D case, the smallest of these values is chosen, ensuring a better validity of the Taylor expansion for the short term evolution of the mean square separation. We then find that the backward Richardson constant is only slightly smaller than the forward constant, $g^{2D}_{bwd}\simeq 6.5$, with a ratio $g^{2D}_{bwd}/g^{2D}_{fwd}\simeq 0.94$ in excellent agreement with simulations by Faber \& Vassilicos who report $g^{2D}_{bwd}/g^{2D}_{fwd} = 0.92\pm0.03$.

\section{Conclusions}
\subsection{Summary of main results}

The iterative ballistic phenomenology approach for turbulent pair separation described in this article presents a simple, robust and intuitive phenomenology which gives a physically sound mechanism accurately describing the overall pair separation scenario, from the short-term ballisitic regime (Batchelor regime) to the celebrated Richardson super-diffusive long-term regime of turbulent relative dispersion. In its simplest formulation, the important physical ingredients in the model are : (i) a short term ballisitic growth of the mean square separation of particle pairs, with a growth rate given by the second-order Eulerian strucutre function of the carrier flow (eq.~\ref{eq:disp1a}), (ii) the usual K41 scaling for the second order Eulerian structure function at inertial scales (given by eq.~\ref{eq:S2}) to account for the scale dependency of the ballisitic growth rate and (iii) a scale dependent duration of the elementary ballisitic process, also given by K41 scalings. The combination of these ingredients leads to a scale-dependent-short-term ballistic process for the mean square separation at each given scale. In this process the scale dependency appears both in the duration of the ballistic regime and in the ballistic growth rate. The existence such a short term ballisitic regime is extremly robust : (i) the quadratic growth of mean square separation, related to the second order strucutre function $S_2$ (eq.~\ref{eq:disp1a}), is a purely kinematic relation (valid beyond the sole frame of turbulence), (ii) the validity of K41 scaling for $S_2$ is a longstanding result of turbulence research (small intermittency corrections could however be included in a refined version of the present model) and (iii) the short term growth of the mean square separation of particle pairs has been shown in previous experimental and numerical studies to follow very precisely the ballisitc relation (\ref{eq:disp1a}) with the usual K41 scaling for $S_2$ and with durations of the ballistic regime also given by K41 scalings~(\cite{bib:bourgoin2006_Science,bib:bitane2012_PRE}). From there, the proposed phenomenology simply consists in propagating iteratively accross scales this scale-dependent-short-term elementary ballistic process. Only two parameters enter into play in this iterative phenomenology : (i) the Kolmogorov constant $C$ which characterizes the ballisitc growth rate at scale $r$ via the second order structure function $S_2(r)=C\epsilon^{2/3}r^{2/3}$ and (ii) the persistence parameter $\alpha$ which characterizes the duration of the ballistic regime at scale $r$, $t(r)=\frac{\alpha C}{2} \epsilon^{-1/3}r^{2/3}$. Note that the persistence parameter $\alpha$ is the only adjustable parameter of this model, as for the Kolmogorov constant we use the well accepted values of the litterature. 

This simple phenomenology reproduces accurately the overall behavior of mean square separation of turbulent relative dispersion. The short term ballisitc Batchelor regime is obviously well described. More interestingly the transition towards the cubic Richardson regime is also well captured, with an explicit connection between the Richardson constant, the Kolmogorov constant and the persistence parameter (eq.~\ref{eq:g}). An interesting aspect of the phenomenology is that the Richardson regime (cubic in time and scale independent) naturally builds itself simply from successive scale dependent ballistic processes. We can also point that although the physical phenomenology is different, this process exhibit interesting analogies with the model proposed by~\cite{bib:goto2004_NJP} for pair dispersion in 2D turbulence.

If the persistence parameter is taken simply equal to 1, as heuristically suggested by~\cite{bib:bitane2012_PRE} simulations (where the transition between the Batchelor and Richardson regime is observed to occur around $t_0=\frac{C}{2} \epsilon^{-1/3}r^{2/3}$), and as done by~\cite{bib:thalabard2014_JFM} in a similar iterative phenomenology, the Richardson constant $g$ is then directly related to the Kolmogorov constant $C$ only. For the case of 3D turbulence ($C=7.7.$) we then find $g=1.0$. Although slightly larger than the well accepted value $g=0.5-0.6$, this is still a reasonalbe estimate considering the absence of any additional adjustable parameter.

Nevertheless, as discussed in section~\ref{sec:ballistic} the assumption $\alpha=1$ for the peristence parameter is not compatible with the ballistic approximation, which requires $\alpha<1$ (and more likely $\alpha\ll1$), and has only a heuristic justification based on the time scale reported in simulations for the Batchelor to Richardson regime transition. The strategy in the present work has therefore been to use the relation~(\ref{eq:g}) between the Richardson constant, the Kolmogorov constant and the persistence parameter to determine the optimal value of $\alpha$, for which the well accepted value of the Richardson constant is recovered. When several values of $\alpha$ satisfy this condition, the smallest value is preferred, to ensure a better validity of the ballistic approximation. For the case of 3D turbulence, the optimal value is $\alpha=0.12$. The predicted evolution of the mean square separation predicted  by the iterative ballistic phenomenology is then in perfect quantitative agreement with numerical simulations by Bitane~\emph{et al.} (fig.~\ref{fig:D2T2}), whatever the initial separation (at inertial scales) for the overall dispersion process. In particular, the collapse of the numerical data at different initial separations is well captured as wall as transition time between the Batchelor and the Richardson (which is found indeed to eventually occur around $t=t_0=\frac{C}{2} \epsilon^{-1/3}r^{2/3}$), and more subtle trends as the slight deceleration of separation growth just before the transition toward the Richardson regime.

Another interesting point predicted by the present phenomenology is that the separation growth can be written in a Richardson-like form at all times, but with a negative virtual time origin~(eq.~\ref{eq:Dk2}), what gives support to the methodology by~\cite{bib:ott2000_JFM} to extract the value of the Richardson constant from relatively short term experimental data of relative dispersion. Besides eq.~(\ref{eq:Dk2}) predicts a simple connection between the virtual time origin, the Richardson constant and the initial separation of pairs, which may be used to reinterpret the virtual time origin determined experimentally.

Finally, a further extension of the model has been proposed to address the question of temporal asymmetry, which is a wellknown feature of turbulent relative dispersion. The pure ballistic phenomenology is time reversible, as it only involves the square of time at all iteration steps. To introduce temporal asymmetry, the Taylor expansion leading to the short term ballistic growth for the mean square separation is pushed to the next order, taking into account the third order corrective term, which involves the crossed velocity-acceleation structure function $S_{au}(\vec{r})=\left<\delta_{\vec{r}}\vec{a}\cdot\delta_{\vec{r}}\vec{u}\right>$. This term is analytically connected, via the Navier-Stokes equation, to the third order structure function of the velocity field, and carries the signature of the energy cascade (its value is $S_{au}(\vec{r})^{3D}=-2\epsilon$ for 3D direct cascade and $S_{au}(\vec{r})^{2D}=+2\epsilon$ for 2D inverse cascade). The present extension of the model has been shown to accurately account for temporal asymmetry reported in experiments and simulations of 3D turbulence (with a direct energy cascade) and in simulations of 2D turbulence (in the inverse cascade regime). In particular, the value of the forward and backward Richardson constants (and their ratio) are quantitatively in agreement with those studies with the persistence parameter as only adjustable parameter. Furthermore, the fact that in 3D turbulence backward dispersion is faster than forward, while the opposite trend is reported for the inverse cascade of 2D turbulence, is explicitly explained in the present phenomenology as the signature of the energy cascade, carried by the thir order term via $S_{au}$, in agreement with the conjecture by~\cite{bib:faber2009_PoF}.

\subsection{Final discussion}


To finish I would like to highlight two important conceptual consequences of the present phenomenology and a few of the several further studies that can be envisaged for the present ballistic phenomenology :

\begin{itemize}

\item An important consequence of the present iterative ballisitc phenomenology concerns the links that it builds between the Lagrangian framework (inherent to the relative dispersion problem) and the Eulerian framework of turbulence. In the present phenomenology the overall process of particle relative dipsersion is, to the leading order, controlled by the Eulerian second order structure function (or equivalently the Eulerian energy spectrum). The iterative ballisitic approach shows indeed that second order Eulerian statistics not only control the short term ballistic separation at each scale, but are also eventually responsible for the emergence of the cubic Richardson regime as a result of the iteration of the ballistic events. Furthermore, the extended version of the phenomenology (including third order terms), relates the temporal asymmetry of relative dispersion to third order Eulerian statistics, and hence to the energy flux accross scales. This justifies the appellation \emph{ballistic cascade phenomenology} chosen for this approach. Furthermore, the iterative phenomenology shows that the temporal asymmetry of turbulent relative dispersion can be quantitatively interpreted simply in terms of inverse \emph{vs} direct energy cascade, as stipulated by~\cite{bib:faber2009_PoF}. Besides, the iterative phenomenology simply connects the short term asymmetry of pair dispersion (given by eq.~(\ref{eq:O3}) or eq.~(\ref{eq:3rdOrder}), and already studied for instance by~\cite{bib:jucha2014_PRL}) to the long term asymmetry (characterized by different values for the forward and the backward Richardson constant). Such a connection still remained to be established, as pointed by Jucha~\emph{et al.}. In particular, the ballisitic cascade phenomnelogy shows that both, short term and long term asymmetry of pair dispersion, can be simply interpreted as a Lagrangian signature of the asymmetry of energy flux accross scales in the Eulerian framework.\\

\item Another interesting remark concerns the fact that the present ballistic phenomenology also offers a new paradigmatic frame to reinterpret what was originally addressed by Richardson as a scale dependent diffusion coefficient (with $K(D)\propto D^{4/3}$, such that the mean square separation at a given scale $D$ grows, locally, linearly in time as $\left<D^2\right>\propto K(D)t$).  Richardson drew this conclusion by a short time, scale by scale, analysis of local diffusion properties over a wide range of phenomena, from diffusion of oxygen intro nitrogen, to the diffusion of cyclones in the atmosphere~(\cite{bib:richardson1926}). It is now accepted that his derivation of the $4/3$rd law was at the same time fortuitous and the result of Richardson's unique intuition~(\cite{bib:sawford2001}). However no robust physical ground support such a scale dependent diffusive scenario. The phenomenology presented here proposes a change of paradigm where the scale dependent \emph{diffusive process} is replaced by a scale dependent \emph{ballistic process}. Besides the ballistic phenomenology also explains why a short time scale by scale analysis can lead to an \emph{apparent} scale dependent diffusive process. The elementary ballistic step given by eq.~(\ref{eq:cascade}) shows that the square separation at each iteration grows by $\left(D_{k+1}^2-D_k^2\right)=S_2(D_k)t_k^{\prime 2}$ what can be rewritten as $[S_2(D_k)t'_k]t'_k=\frac{\alpha C}{2} \epsilon^{1/3}D_k^{4/3} t'_k$. The short term growth of the mean square separation, can therefore be equivalently interpreted as a scale dependent normally diffusive process with a diffusion coefficient $K(D)=\frac{\alpha C}{2} \epsilon^{1/3}D^{4/3}$, precisely following Richardson's ``4/3rd'' law.\\

\item As discussed in section~\ref{sec:relevance}, an important approximation of the iterative scheme, concerns the fact that the estimation of the duration $t'_k(D_k)$ of each iteration step and of the ballistics growth rate is done based only on the mean square separation $D_k^2$, but does not take into account the fact that separations spread over a whole statistical distribution. In spite of this approximation, the results presented here show \emph{a posteriori} that the model behaves extremely well to predict the evolution of the mean square separation. A first test to probe \emph{a priori} the relevance of this approximation can be done using the measured (experimentally or numerically) statistics of particles relative dispersion, in order to estimate the evolution of the width of the statistical distribution of the square separation (which can be estimated from the flatness of the distribution). The approximation can be assumed to be valid as long as this width remains small compared to the mean square separation itself. Beyond this simple verification, it is also possible to imagine to implement a similar iterative approach for higher order moments (and not only for the mean square separation), what should then involve higher order velocity strucutre functions, and hence also address intermittency corrections issues related to the relative dispersion problem. These possible extensions will be investigated in future studies.\\ 

\item Many situations of practical interest for turbulent dispersion issues concern non-homogeneous flows. This is the case for instance for the dispersion of pollutants in the atmospheric boundary layer. Extending the present phenomenology to the case of non-homogeneous turbulence is therefore also of relevant interest. This would involve several important steps forward. One aspect concerns taking into account inhomegeneities of the velocity field, to account for instance for the dependency of the second order structure function not only on the pair separation, but also on the pair position. Another aspect concerns the approximation, also discussed in section~\ref{sec:relevance}, which consists in neglecting the term $<\vec{D}_0\cdot\delta_r \vec{u}>$, in order to apply the elementary ballisitic growth to $<D^2>-<D_0^2>$ instead of $<(\vec{D}-\vec{D}_0)^2>$. This approximation has been shown to fail, at very short times, in non-homogeneous experimental flows~(\cite{bib:ouellette2006b_NJP}). It can be shown that keeping this term, adds a short term linear contribution to the growth of the mean square separation compared to the ballisitc scenario considered here. This will tend to add a short term normally diffusive-like contribution. The extension to non-homogeneous situations will therefore probably require to investigate the relative importance of this linear contribution compared to the ballistic contribution.\\

\item The physical origin of the persistence parameter, which controls the statistical duration of each ballistic step remains to be clarified. The analogy with the model by~\cite{bib:goto2004_NJP} suggests a possible interpretation in the case of 2D, in terms of the probability to encounter hyperbolic stagnation points. More generally, it would be interesting to investigate possible connections with the idea of delay times, recently proposed by~\cite{bib:rast2011_PRL} to possibly play a leading role in turbulent pair dispersion.

\item The present ballistic phenomenology goes beyond the sole frame of turbulent dispersion, as it shows that super-diffusion is not an exclusive feature of turbulence. Pairs separating super-diffusively can be simply driven by pure ballistic processes, as soon as the timing and/or the ballistic growth rate is controlled scale by scale. This shows the way towards new possible strategies to enhance dispersion in non turbulent flows, where efficient mixing is difficult to achieve. This is the case for instance in micro-fluidic devices, where the ability to control and tune phoretic mechanisms (as diffusiphoresis~(\cite{bib:abecassis2008_NatMat}), thermophoresis, chemotaxis, photophoresis, etc.) could be efficiently used to drive a ballistic cascade of the motion of colloidal particles, macro-molecules, bacteria, etc., in order to promote their super-dispersion.
 
\end{itemize}


\acknowledgments{I would like to acknowledge Yves Gagne and Fr\'ed\'eric Moisy for useful discussions. I am also thankful to fruitful suggestions from the anonymous referees. This work is part of the International Collaboration for Turbulence Research. It has received financial support from the French National Research Program ANR-12-BS09-0011 ÔTEC2Õ.}

\bibliographystyle{jfm}
\bibliography{main}
\end{document}